# A Feature Engineering Approach for Literary and Colloquial Tamil Speech Classification using 1D-CNN

M. Nanmalar[1], S. Johanan Joysingh[1], P. Vijayalakshmi[1], and T. Nagarajan[2]

[1]Department of ECE, Sri Sivasubramaniya Nadar College of Engineering, Chennai, India.
[2]Department of CSE, Shiv Nadar University Chennai, India.

# 1 Abstract

In ideal human computer interaction (HCI), the colloquial form of the language would be preferred by most users, since it is a form used in their day-to-day conversations. However, if that language is rich and abundant in literature, and has more than 2000 years of history, such as the Indian language Tamil, there also comes an undeniable necessity to preserve it. By embracing the new and preserving the old, both service to the common man (practicality) and service to the language itself (conservation) is rendered. This can enable sustainable preservation and growth of the language, culture, tradition and literature. In light of this, computers must have the ability to accept, process and converse in both forms of the language, as required. To address this, it is first necessary to identify the form of the input speech, here literary or colloquial Tamil. Such a front-end system must consist of a simple, effective, and lightweight classifier that is trained on a few effective features that are capable of capturing the underlying patterns of the speech signal. To accomplish this, in the current work, a one-dimensional convolutional neural network (1D-CNN) that learns the envelope of features across time, is proposed. The network is trained on a select number of handcrafted features initially, and then on Mel frequency cepstral coefficients (MFCC) for comparison. The handcrafted features were intuitively selected to address various aspects of speech such as the spectral and temporal characteristics, prosody and voice quality. The relevance of each feature, in its ability to distinguish the two classes, is initially analyzed by considering ten parallel utterances and observing the trend of each feature with respect to time. The proposed 1D-CNN, trained using the handcrafted features, offers an F1 score of 0.9803, while that trained on the MFCC offers an F1 score of 0.9895. Since both sets of features are effective, feature combination is explored. However, to quantitatively measure and rank the contribution of each handcrafted feature, a feature ablation study is first carried out, which aids in eliminating features that are least useful. When the best ranked handcrafted features, from the feature ablation study, are combined with the MFCC, they offer the best results with an F1 score of 0.9946. From the analysis and experimental results it can be concluded that: (i) distinguishing information is present in the envelope of the proposed features, (ii) the proposed 1D-CNN is able to learn the trends in these features over time, and (iii) the handcrafted features complement the information offered by MFCC.

# 2 Introduction

More than 80 million people in the world speak Tamil, but the form of Tamil they speak now is different from what their ancestors spoke a century ago. This is due to the inevitable evolution that every language goes through. One interesting aspect about Tamil is that both a literary and colloquial form co-exists in different contexts. The archaic and formal form of the language is called literary Tamil (LT), while the contemporary and casual form of the language is called colloquial Tamil (CT) in the current work. CT was built on top of LT, hence there is a considerable level of similarity between them. The differences between these two forms of Tamil can be majorly seen in the lexical and acoustic characteristics. Lexical differences can occur in the sub-word, word, and phrase level, while the acoustic differences can occur in style, effort, speech rate, etc.

In [1], the acoustic differences between read (prepared) speech and spontaneous speech are analysed, and their effects on a dialect identification task is studied. With reference to the current work, read speech is analogous to literary Tamil, and spontaneous speech to colloquial Tamil. This is due to the fact that literary

Tamil is more standardized, and the text is parsed and spoken as such from its written form, similar to read speech. While on the other hand, colloquial Tamil is mainly a spoken language with no real standards in its written form, hence being more similar to spontaneous speech. It can be stated that speakers of CT liberate themselves from the standards of written LT to write CT.

In [2], it is shown that the spectral (MFCC) space shrinks for spontaneous speech, when compared to read speech. This reduction is quantitatively provided using a metric known as the reduction ratio. This ratio is computed for each phoneme, by considering the Euclidean distance between the mean vector that corresponds to the set of all phonemes, and the mean vector of that phoneme, for spontaneous and read speech. It is shown that the reduction ratio is lower for more spontaneous speech, meaning that the characteristics of the various phones in the dialect are more close to each other, making phone recognition harder. Hence, it is important that read and spontaneous speech are addressed specifically.

In current automatic speech recognition systems, colloquial Tamil is not addressed as well as literary Tamil. This is true in the case of text-to-speech synthesizers and machine translation systems as well. The fact that colloquial Tamil requires specific attention can be understood from the experiments conducted in [1]. In [1], when an Arabic dialect identification (DID) system trained on read speech was tested on read speech, the error rate was determined to be 17%. Similarly, when spontaneous speech was tested on a system trained on spontaneous speech, the error rate was 20%. But when a system trained on read speech was tested on spontaneous speech, the error rate rises to 37%, thus showing the characteristic differences between the two forms of speech. Again, in [3], the performance of read and spontaneous speech are compared with respect to a prosody-based language identification (LID) system. From the experiments in [3], it can be seen that the LID system performs better for read speech, but struggles to address the complexity of the prosody inherent in spontaneous speech, adding to the fact that colloquial Tamil requires specific attention.

In light of the evidence presented so far, it can be seen that in the context of Tamil, a literary-colloquial Tamil identification (LCTID) system is an essential front-end system. It should also be noted that there has been no particular focus on the research of these two dialects of Tamil speech so far, which is evident from the detailed literature survey table presented in [4]. In line with this, in our previous work [5] a few implicit and explicit LCTID systems were explored. These include, Gaussian mixture models, parallel phone recognition, parallel-large vocabulary continuous speech recognition, unified phone recognition, and 1DCNN-based dialect identification, using MFCC features on a newly developed in-house corpus. Of the implicit and explicit LCTID systems, the implicit varieties are more practical due to their text-independent nature. This is especially accented by the fact that the written form of CT does not have any particular standard, equivalent to LT. Besides, an ideal LCTID system must be accurate and computationally efficient, since it is a front-end system. Considering these aspects, a 1DCNN-based LCTID system is deemed to be a suitable fit, and hence is pursued further in the current work.

The current work investigates the potential of 1D-CNN in learning intermediate and discriminating features from handcrafted features extracted from literary and colloquial Tamil speech. The existing work related to LCTID, features for dialect and language identification, and significance of 1DCNN are discussed in Section 3. Following this, some of the significant acoustic differences between literary and colloquial Tamil are discussed in Section 4, which will form the basis for the proposed handcrafted features. In Section 5, the suitability of these handcrafted features for LCTID, is discussed through an analysis of the trend of these features for a set of parallel utterances. The experimental setup, which include the architecture of the 1DCNN, the training methodology, and the details of the dataset are provided in Section 6. In Section 7, the experimental results and observations are discussed. Following which, a comparison of MFCC features with the proposed handcrafted features, along with a feature ablation study of the handcrafted features is provided. Furthermore, a novel feature set combining the best ranked handcrafted features and the MFCC is proposed, and the results are interpreted.

## 3 Existing Related Works

A summary of handcrafted features for dialect and language identification, followed by a discussion on the significance of 1DCNN in learning time-series data, based on existing literature, is provided in this section.

### 3.1 Features For Dialect Classification

Every dialect has unique characteristics, and it is important to represent these unique characteristics as features. Prosody is proposed by many in literature as a feature for dialect classification. Especially in the case of overlapping phonetics between the dialects, prosody can offer more discriminating information than acoustic-phonetic or phonotactic features. The handcrafted featureset considered in the current work include prosodic, voice quality, spectral, and temporal features. Apart from the prosodic features, voice quality features can be said to contain prosodic information, since it manifests micro-prosodic variations. The temporal feature, zero crossing rate, is also considered as part of a prosodic featureset by some, as in [6]. Hence, the literature below majorly focuses on prosodic and spectral features.

### 3.1.1 Perceptual Analysis of Prosodic Features

A study to evaluate whether prosodic patterns can be considered as reliable acoustic cues for the discrimination of Arabic dialects is presented in [7]. This is carried out through a perceptual experiment, where listeners who have access to only prosodic cues such as fundamental frequency, amplitude, and some rhythmic characteristics of the original signal were able to identify the dialects successfully. Another similar study is presented in [8], where the authors aim to find if prosody alone can act as a significant feature for the identification of the regional origin of Swiss-German speakers. The study considers four Swiss-German dialects. The experiment involves 70 participants listening to a delexicalized speech signal, band-pass filtered from 250 Hz to 7000 Hz. It is observed that despite this filtering, 3 of the 4 dialects were successfully recognized by the participants. Furthermore, it is observed that the identification rates are considerably higher for dialects which present distinct prosodic features.

### 3.1.2 Classifiers Using Prosodic Features Alone

In line with the observations and conclusions of the perceptual analysis, various language and dialect identification systems have been proposed in literature which utilize only prosodic features. In [9], a spoken Algerian-Arabic dialect idenfication system using prosodic features alone, is proposed. Here, intonation features such as $f_0$, and rhythm features using the Ramus model, Dellwo model, and Grave model are proposed. In [10] a language identification system is developed using prosodic features alone. The prosodic features considered here are either the first difference of $F_0$, or the first difference of the bandpass filtered amplitude envelope. In [11] a language identification system based on $F_0$, energy and duration features is developed for the identification of seven languages. In [3], $F_0$ and rhythm parameters are used for language identification for both read and spontaneous speech. In [12], a dialect classifier for the Assamese language is presented, wherein two different prosodic features of vowel sounds, one using the first four formants of a particular vowel sound, and the other using individual formants of all the vowel sounds, are explored as features for classification.

### 3.1.3 Classifiers Using Prosodic and Spectral Features

Even though only prosodic features were considered in many of the previously mentioned proposals, there are a handful of proposals where both prosodic and spectral features are used. In [13], nine dialects of British English are classified using a set of prosodic features such as pitch, energy and duration, and a set of spectral features such as MFCC, shifted delta coefficients (SDC), spectral flux, and spectral entropy. In [14], five Hindi dialects are classified using prosodic features such as pitch, energy and duration, and MFCC. In [15], five kannada dialects are classified using prosodic features such as pitch, energy and loudness, and spectral features such as MFCC, energy entropy, spectral centroid, spectral spread, and spectral flux. Similarly in [15] prosodic features derived from different dimensions of $F_0$ with its delta features, along with MFCC features are proposed for the classification of two Ao dialects. In [16] three chinese dialects are classified using pitch flux (prosodic) and MFCC (spectral) features.

### 3.1.4 Feature Selection

In the previous sections, the perceptual importance of prosodic features was first reviewed. Following which, classifiers which utilize the prosodic features alone, followed by those that use both prosodic and spectral features were reviewed. Some concluding remarks follow. In [17], since language recognition performance was better when utilizing spectral information compared to prosodic information, authors suggest the combination of both spectral and prosodic features. An interesting aspect of feature selection for classification of language, dialects, and speakers is specified in [18]. Here, it is proposed that, since language and speaker recognition systems based on spectral features perform well in favorable acoustic conditions [19], and since prosodic features derived from pitch, energy, and duration are relatively less affected by channel variations and noise [20], a system based on both spectral and prosodic features would provide the necessary robustness to these recognition systems. From this review of literature it is evident that the combination of prosodic and spectral features is a valid route for the classification of language or dialects.

## 3.2 Significance of 1D-CNN

The advantages of CNN in general are that, (i) it extracts intermediate features before classification, (ii) it is less complex when compared to deep neural networks that can produce the same accuracy, and (iii) it is tolerant to small transformations such as distortion, scaling, and skewing. The 1D-CNN was first proposed in [21] and [22]. The main advantages of this new architecture were the following.

1. Ability to learn directly from one dimensional data.
2. Ability to learn complex and challenging tasks even with a shallow architecture.
3. Compact architecture with only a few hidden layers.
4. Much reduced computational complexity when compared to 2D-CNN, and the possibility of being trained on CPUs, instead of GPUs.
5. Ability to converge on small datasets.
6. Real-time application on handheld devices.

In [23] 1D-CNN is proposed for the classification of shouted and normal speech. Here, the 1D-CNN learns the harmonic structure from the magnitude spectrum of speech frames. In [4], the accent classification of three major Nigerian indigenous languages is carried out using a 1D-CNN and LSTM-based network. From these discussions, it is evident that the 1D-CNN is well suited for the current work.

# 4 Acoustic Differences Between LT and CT

In order to find the right features for a classification task, the distinct characteristics between the classes must be known. Since CT evolved from LT, they are quite similar in many aspects, however there are some notable characteristic differences, which are discussed in this section. The generic differences in speaking style between LT and CT are summarized in Table 1. These differenes are usually reflected in the following

Table 1: Differences in speaking style between LT and CT

|  | **LT** | **CT** |
| --- | --- | --- |
| Pronunciation | Formal | Casual |
| Flexibility | Strict | Linient |
| Standards | Yes | No |
| Effort | High | Low |
| Fillers | Not Common | Common |
| English | Absent | Present |
| Other Languages | Absent | Present |
| Pauses | Language Defined | User Defined |
| Assimilation | Language Defined | User Defined |

manner.

## 4.1 Pronunciation Effort

In CT, the phonemes that require more effort to pronounce are usually avoided. These include phonemes such as /au/, /ai/, /zh/ and /r/. These phonemes add uniqueness to literary Tamil. Besides, Tamil and Malayalam are amongst the few languages in the world which have the phoneme /zh/. Some examples of how these phonemes are modified in CT is shown below,

- /au/: awvaiyaar (ஔவையார்) ⟶ avvayaar (அவ்வயார்)
- /ai/: aindu (ஐந்து) ⟶ anju (அஞ்சு)
- /zh/: Tamizh (தமிழ்) ⟶ Tamil (தமில்)
- /r/: sirappu (சிறப்பு) ⟶ sirappu (சிரப்பு)

Furthermore, phonemes that are very similar in pronunciation, and require effort to distinguish, are usually fused into a single phoneme. For example, the phonemes, /l/, /lx/ and /zh/, are all pronounced as /l/, and the phonemes /r/ and /rr/ are both pronounced as /r/.

## 4.2 Word-Ending

Unlike LT, almost all CT words end in a vowel. In other words, the consonant ending in LT is converted to vowel ending in CT. This is usually by, (i) the addition of a vowel, or (ii) the deletion of the consonant, at the end of the word. Some examples of addition are:

- kan (கண்) ⟶ kannu (கண்ணு)
- pal (பல்) ⟶ pallu (பல்லு)

Some examples of deletion are:

- vandhaal (வந்தாள்) ⟶ vandhaa (வந்தா)
- seidhaal (செய்தாள்) ⟶ senjaa (செஞ்சா)

A consequence of vowel endings is that, connected speech is produced, which leads to reduced phrase breaks. Reduced phrase breaks, together with reduced effort to produce phonemes, encourages higher speech rate in CT, leading to shorter utterances. The reduction in duration can be observed in Figures 1 through 10. Hence one could say that the speakers of CT attempt to end words in a vowel for the sake of ease of speech in everyday conversations.

## 4.3 Nasalized Vowel

In the context of Tamil speech, the coarticulation effect of the nasal consonant preceding, and/or succeeding, a vowel, produces a nasalized vowel. The nasalization can either be accidental or distinctive, based on the degree to which the vowel is nasalized [24]. For example, considering the two phrases 'she danced (ஆடினா)' and 'he danced (ஆடினா(ன்))', the latter has distinctive nasalization. More specifically, distinctive nasalization occurs when the words of more than one syllable, ending in a vowel plus the nasal consonants /m/ and /n/, change to nasalized vowels, and the nasal segment is deleted. Only this type of nasalization is considered in the current work. It can be given by the following formulation,

$$xx + V + NC \to xx + NV, \qquad (1)$$

where $xx$ represents one or more syllables preceding the last syllable of a word, $V$ is the vowel of the last syllable, $NC$ is the nasal consonant of the last syllable, and $NV$ is the nasalized vowel. This characteristic is very unique to CT. Some examples of vowel nasalization follows:

- maram (மரம்) → maro(m) (மரோ(ம்))
- varam (வரம்) → varo(m) (வரோ(ம்))
- pazham (பழம்) → pazho(m) (பழோ(ம்))

There are no representation for nasalized vowels in the orthographic form in the language Tamil. However, for ease of understanding, in the current work, it is represented as the nasal consonant within parenthesis, as shown in the examples above.

## 4.4 Code Switching

Code switching is the process where the vocabulary of a primary language gets temporarily switched with that of a secondary language, which happens mostly when people live near the border of two states. In the case of CT, code switching mainly occurs with respect to English more than other Indian languages, and happens quite casually. Due to this, new phones are added to the phoneset of CT.

The first phoneme in LT will predominantly not be a voiced consonant. However in CT, there is no such characteristic (no strict standard) because of the loan words borrowed from English, and sometimes from other Dravidian languages (such as Telugu and Malayalam), and Indo-Aryan languages. This characteristic, similar to the nasalized vowel, is not represented orthographically. More interestingly, Tamil orthography does not offer the freedom for the representation of voiced consonants. For example, the words "fashion" and "ball", when written in Tamil, will be pronounced as "pason (பாசன்)" and "paal (பால்)".

## 4.5 Unvoiced Stops

One of the characteristic features of LT is the occurrence of unvoiced stops at the end of the words, except the last word. This causes minor breaks between words. This grammatical feature called "sandhi" is mainly present in LT for the purpose of beautification, and is not appreciated in CT, for the sake of effortless speech. Some examples follow,

- seyyach chonnaan (செய்யச் சொன்னான்) ⟶ seyya sonnaa(n) (செய்ய சொன்னா(ன்))
- chellap peran (செல்லப் பேரன்) ⟶ cella pera(n) (செல்ல பேர(ன்))

Table 2: Parallel sentences 1-5, used in the analysis of the feature contours in LT and CT.

|   | English (Literal Translation) | Literary Tamil | Colloquial Tamil |
|---|---|---|---|
| 1 | What are you doing | /nd ii/ /e n n a/ /c ey g i rx aa y/ (நீ என்ன செய்கிறாய்) | /nd ii/ /e n n a/ /c e y y i rx a/ (நீ என்ன செய்யிற) |
| 2 | Twice a week, we have leave. | /w aa r a m/ /i r u/ /m u rx ai/ /e ng g a lx u k k u/ /w i dx u m u rx ai/ (வாரம் இரு முறை எங்களுக்கு விடுமுறை) | /w aa r o (m)/ /r e nx dx u/ /d a dx a w a/ /e ng g a lx u k k u/ /l ii w u/ (வாரொ(ம்) ரெண்டு தடவ எங்களுக்கு லீவு) |
| 3 | The sight of a lion, makes me fear. | /c i ng g a t t ai p/ /p aa r t t aa l/ /e n a k k u/ /b a y a m/ (சிங்கத்தை பார்த்தால் எனக்கு பயம்) | /c i ng g a t t a/ /p aa t t aa/ /e n a k k u/ /b a y o (m)/ (சிங்கத்த பாத்தா எனக்கு பயொ(ம்)) |
| 4 | The tiger was wailing continuously. | /p u l i/ /k a t t i k o nx dx ee/ /i r u nd d a d u/ (புலி கத்திக்கொண்டே இருந்தது) | /p u l i/ /k a t t i tx tx ee/ /i r u nd d u c c u/ (புலி கத்திட்டே இருந்துச்சு) |
| 5 | I would become very sad. | /nd aa n/ /m i g a w u m/ /c oo r nd d u/ /p oo y/ /w i dx u w ee n/ (நான் மிகவும் சோர்ந்து போய் விடுவேன்) | /nd aa (n)/ /r o m b a/ /c oo nd d u/ /p oo y dx u w ee (n)/ (நா(ன்) ரொம்ப சோந்து போயிருவே(ன்)) |

## 5 Feature Set

The legacy LID and DID systems such as [25] are acoustic-phonetic and phonotactic-based approaches. They offered very good results and formed the foundation of many LID and DID approaches that were proposed later. Another avenue of research focused on developing handcrafted features with the sole purpose of crafting features based on the style and uniqueness of the language or dialect, without confining itself to acoustic-phonetic or phonotactic features. Over the years, the length of the segment considered for feature extraction has shifted from the phone, to the syllable, to the psuedo-syllable, to a sequence of syllables, and finally to suprasegments. Yet in recent times, hand-crafted and learnable framewise features, combined with time-dependency, are being considered in the state-of-the-art systems for better performance [26]. In this work, ten framewise handcrafted features are proposed and are intuitively chosen that they may work along with the feature engineering capabilities of the 1D-CNN, which capture their numerous varying patterns over time. Out of the ten hand-crafted features proposed, three are prosodic features, four are voice quality features, two are spectral features, and one is a temporal feature. The proposed features are the following:

- Prosodic features – describes the tonality, melody and dynamics of speech.
  - Fundamental frequency
  - Energy
  - Voicing Probability
- Voice quality features — describes micro-prosodic variations.
  - Jitter
  - Derivative of Jitter
  - Shimmer
  - Harmonic-to-Noise Ratio (HNR)

Table 3: Parallel sentences 6-10, used in the analysis of the feature contours in LT and CT.

|   | English (Literal Translation) | Literary Tamil | Colloquial Tamil |
|---|---|---|---|
| 6 | That crowd would come here. | /a nd d a/ /k uu tx tx a m/ /i ng g ee/ /w a r u m/ (அந்த கூட்டம் இங்கே வரும்) | /a nd d a/ /k uu tx tx o (m)/ /i ng g a/ /w a r u (m)/ (அந்த கூட்டொ(ம்) இங்க வரு(ம்)) |
| 7 | You would get sweet pongal, tamarind rice, curd rice, and sambhar rice in the perumal temple. | /p e r u m aa lx/ /k oo w i l i l/ /c a r k k a r ai p/ /p o ng g a l/ /p u lx i/ /c aa d a m/ /t a y i r/ /c aa d a m/ /c aa m b aa r/ /c aa d a m/ /e l l aa m/ /k i dx ai k k u m/ (பெருமாள் கோவிலில் சர்க்கரை பொங்கல் புளி சாதம் தயிர் சாதம் சாம்பார் சாதம் எல்லாம் கிடைக்கும்) | /p e r u m aa lx/ /k oo w i l l a/ /c a k k a r a/ /p o ng g a l/ /p u lx i/ /c aa d o (m)/ /t a y i r/ /c aa d o (m)/ /c aa m b aa r/ /c aa d o (m)/ /e l l aa (m)/ /k e tx e k k u (m)/ (பெருமாள் கோவில்ல சக்கர பொங்கல் புளி சாதொ(ம்) தயிர் சாதொ(ம்) சாம்பார் சாதொ(ம்) எல்லா(ம்) கெடைக்கு(ம்)) |
| 8 | My elder sister came buying samosas. | /e n/ /a k k aa/ /c a m oo c aa/ /w aa ng g i k k o nx dx u/ /w a nd d aa lx/ (என் அக்கா சமோசா வாங்கிக்கொண்டு வந்தாள்) | /ee (n)/ /a k k aa/ /c a m oo c aa/ /w aa ng g ii tx tx u/ /w a nd d aa/ (ஏ(ன்) அக்கா சமோசா வாங்கிட்டுவந்தாள்) |
| 9 | My leg pained a lot. | /e n a k k u/ /m i g a w u m/ /k aa l/ /w a l i t t a d u/ (எனக்கு மிகவும் கால் வலித்தது) | /e n a k k u/ /r o m b a/ /k aa l/ /w a l i c c u d u/ (எனக்கு ரொம்ப கால் வலிச்சுது) |
| 10 | My heart is usually good. | /p o t u w aa g a/ /e n/ /m a n a c u/ /t a ng g a m/ (பொதுவாக என் மனசு தங்கம்) | /p o d u w aa/ /ee (n)/ /m a n a c u/ /t a ng g o (m)/ (பொதுவா ஏ(ன்) மனசு தங்கொ(ம்)) |

- Spectral features — describes the spectral properties of the speech. Cepstral features are not included in this set.
  - Spectral Flux
  - Psychoacoustic Sharpness
- Temporal feature, Zero Crossing Rate — describes the frequency dynamics of the signal.

In order to evaluate the appropriateness of these features, the following brief analysis is carried out. Here, ten parallel LT and CT sentences are formulated (refer Tables 2, and 3), where parallel sentences imply that they both have the same meaning. These sentences are then recorded from five speakers twice, for a total of 200 utterances. The ten proposed hand-crafted features are then extracted from these sentences, and the trend of these features are plotted across time (feature-trend plots) and studied. A brief description of these hand-crafted features, along with a discussion of their feature-trend plots is provided below. These plots provide tangible evidence of the distinguishing patterns that these features offer.

## 5.1 Prosodic Features

Although speech can be characterized by its source and its system characteristics, real speech is characterized by much more properties which provides its naturalness. Similarly, sound units of real speech do not have static characteristics, but rather is characterized by properties such as the rise and fall of the tone (intonation), expansion and compression (duration), and accentuation. The sound units of a language make up what is being said, while its prosody make up the way it is being said, and is often described as a musical quality of speech. Without prosody, speech can sound monotonous and boring, hence it makes sense why it is sought after in human computer interfaces to address naturalness. Prosody also captures the style of speech, which can be used to identify intentions, attitude and affect. Hence, we propose that prosody related acoustic correlates would be greatly helpful in capturing the style and acoustic differences between LT and CT as mentioned in Section. 4. A discussion of the chosen prosodic features follows.

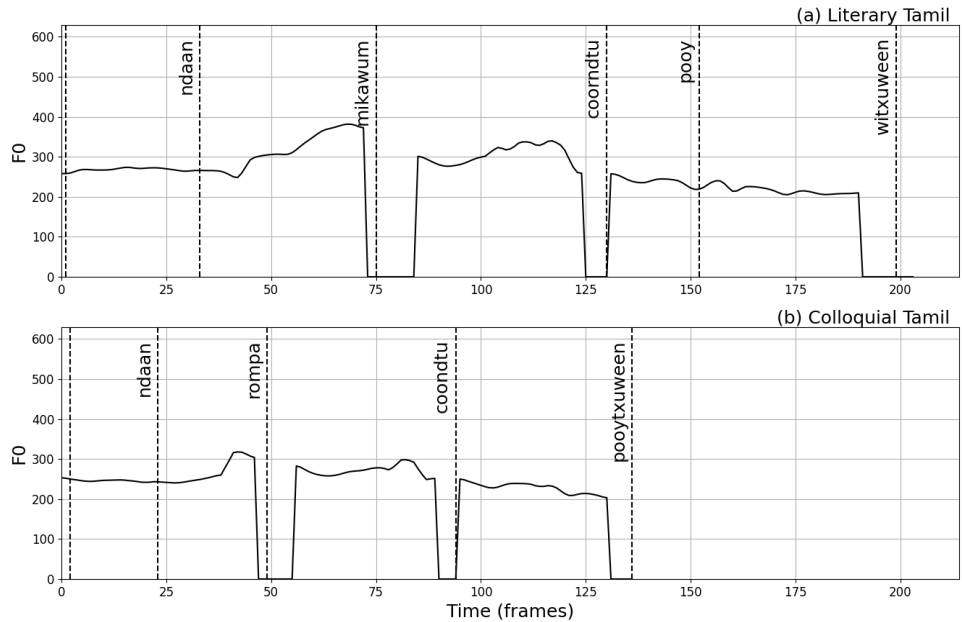

(a) Sentence 5

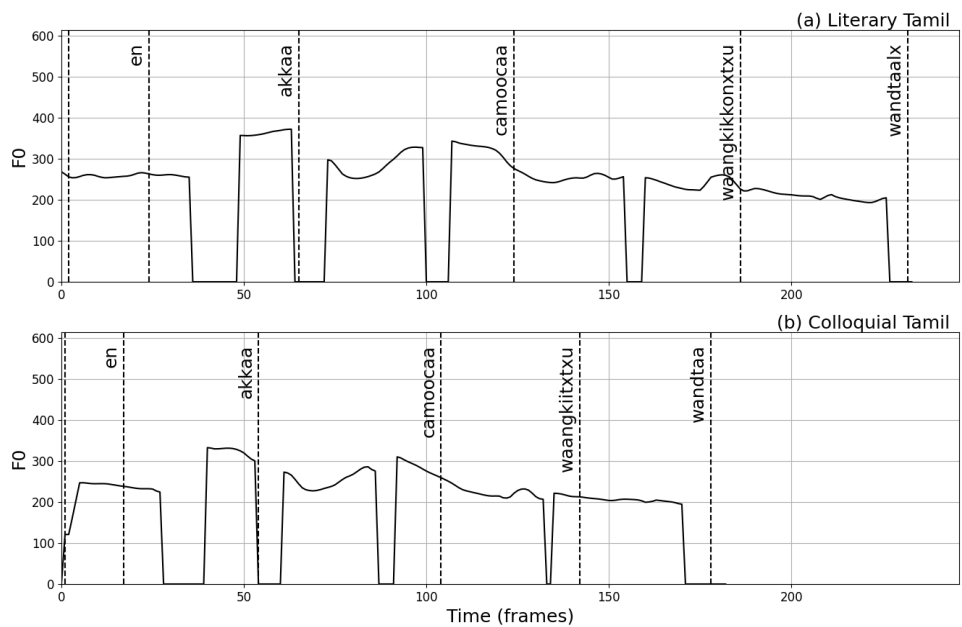

(b) Sentence 8

Figure 1: $F_0$ contour of a pair of parallel LT and CT sentences

### 5.1.1 Fundamental Frequency

A straight forward definition of pitch is that it corresponds to the frequency of a tone. However, it is a perceptual term [27], and is usually defined as the 'perceived tonality'. The physical correlate of pitch is the fundamental frequency ($F_0$), which is the lowest frequency in the harmonic series of a tone. In terms of speech, it corresponds to the rate of vibration (or frequency) of the vocal folds. Comparisons in [28] show that languages differ greatly in this aspect, and so do dialects. An exhaustive comparison of the intonation patterns in twenty languages, which can be found in [29], conveys the same. Fundamental frequency as a feature is prevalent in all existing works based on prosody for LID and DID systems [3, 7, 8, 9, 10, 11, 13, 14, 15, 16, 30, 31, 32]. The reason for its prevalence is due to the fact that it is considered less sensitive to channel distortions and additive noise [18, 33, 34], and hence is robust. Speaker verification using $F_0$ features in [35] show that $F_0$ features improve verification performance at low SNR. Further evidence of this robustness can be found in [34], where it is suggested that people with cochlear implants benefit from the fine structure cue provided by fundamental frequency ($F_0$), in noisy environments. Contours of $F_0$ are also deemed to be language dependent, and speaker independent, as conveyed in [18], where $F_0$ contours of Mandarin and Farsi spoken by multiple speakers are compared.

In the current work $F_0$ is computed using the subharmonic-summed spectrum (SHS) method [36], which is considered a robust pitch estimation technique. It is based on human perception principles, and relies on the harmonic structure of the signal to identify the pitch, and hence works even when the fundamental frequency is absent (missing fundamental [37]). It works by first summing shifted versions of a log-scaled spectrum, which is auditory weighted using an arctangent function. The maximum of a set of $N_p$ pitch candidates (peak magnitudes of local maximas), derived from this subharmonic summed spectrum, using a greedy peak picking algorithm, is the determined pitch.

Figure 1 shows the $F_0$ contour of a pair of parallel LT and CT sentences, which are sentences 5 and 8 in Tables 2 and 3 respectively. It can be observed from the Figure that, the $F_0$ contour of CT is smoother when compared to LT. The variations in the $F_0$ contour produced due to the expressiveness of literal speech, is absent in colloquial speech which prioritizes ease of speech. Specifically, consider the last two words in LT, and the last word in CT, in both examples. Whereas there is a rise and fall in pitch in each syllable of the former, the latter has a relatively flat pitch.

### 5.1.2 Energy

Energy is considered one of the basic, yet powerful audio descriptors [38]. Pitch and energy are popular in LID and DID tasks [7, 11, 13, 14, 30, 32], apart from tasks such as VAD and emotion recognition.

Energy quantifies the stress associated with speech. Since stress variations are characteristic of a language or dialect [39], energy contours are suitable features for LID and DID tasks. In the current work, the energy in each frame is computed as the sum of the squared amplitudes in that frame, assuming no DC offset. It is given by,

$$E = \sum_{n=0}^{N-1} x^2(n), \qquad (2)$$

where $x(n)$ are the samples in a frame of size $N$, and $E$ is the computed energy.

Figure 2 shows the energy contour of a pair of parallel LT and CT sentences, which are sentences 1 and 5 in Table 2. It can be observed from the Figure that a similar trend to that of $F_0$ exists for energy as well. That is, the energy contour of CT is smoother than that of LT. The contour of LT contains many minor peaks apart from the major peaks which are present in both LT and CT. This can be attributed to the fact that in LT, each syllable is individually stressed, while in CT, to aid ease of speech, they are connected.

### 5.1.3 Voicing Probability

Voicing probability indicates the measure of closeness of a given signal with respect to either a harmonic signal, or a noise-like signal. It produces higher values for harmonic signals, and lower values for noise-like signals [40]. In the current work, it is derived using the SHS algorithm, explained in section. 5.1.1. It can be a good alternative to zero crossing rate, which is not dependable at low SNR conditions. The voicing probability associated with each pitch candidate $i$ is given by,

$$p_{v,i} = 1 - \frac{\mu_H}{X_i} \qquad (3)$$

where $\mu_H$ is the mean of the subharmonic-summed spectrum, $X_i$ are the amplitudes of the pitch candidates. The final voicing probability is that which is derived for the pitch candidate with the highest amplitude.

Figure 3 shows the voicing probability contour of a pair of parallel LT and CT sentences, which are sentences 5 and 10 in Tables 2 and 3 respectively. Similar to $F_0$, the voicing probability rises and falls along with the utterance of each syllable, due to the presence of voiced and unvoiced regions in a syllable. While for CT, the rate of voicing and unvoicing is lower, for LT it is higher. In CT, the voicing probability contour does not vary as much as that of LT, due to the deletion of phones in the parallel word. This can be seen in the words, "sorndhu poividuven (சோர்ந்து போய்விடுவேன்)"" in LT, and "sondhu poiruve(n) (சோந்து போயிருவே(ன்))" in CT, where the phones that require effort are either avoided or replaced. In Figure 3(b), the syllable 'gam' and 'go(m)' from the ending of words 'thangam (தங்கம்)' and 'thango(m) (தங்கொ(ம்))' which are LT and CT words respectively. It can be observed that for the latter, the peak is more pronounced, than the former. This is due to the presence of a consonant at the end of the word in LT, while for CT the nasal consonant combines with the preceding vowel and presents itself as a nasalized vowel, resulting in higher voicing probability. Similar patterns can be observed in the word pairs 'bayam (பயம்) - bayo(m) (பயொ(ம்))', and 'varum (வரும்) - varu(m) (வரு(ம்))' as well.

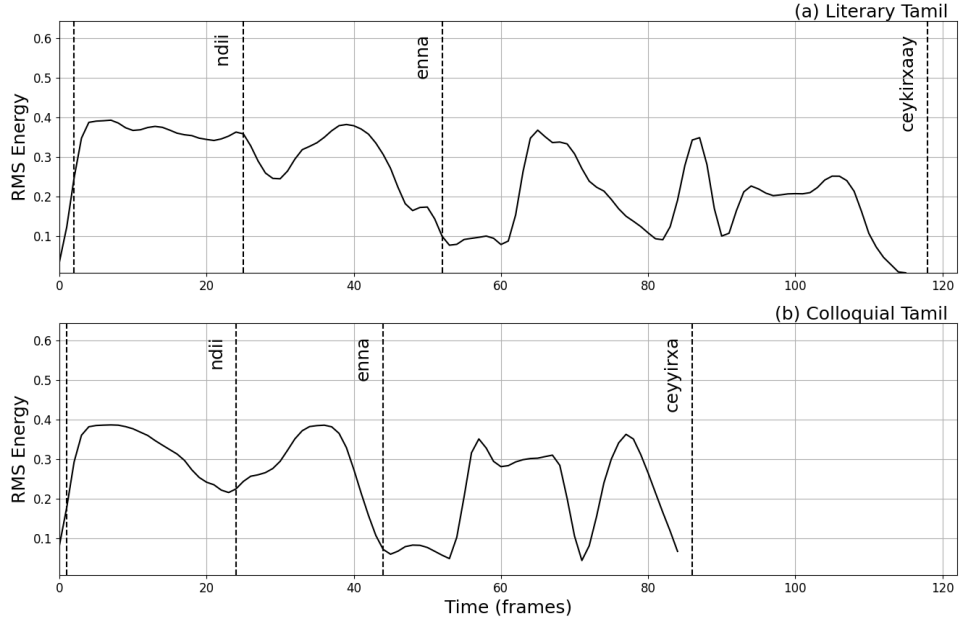

(a) Sentence 1

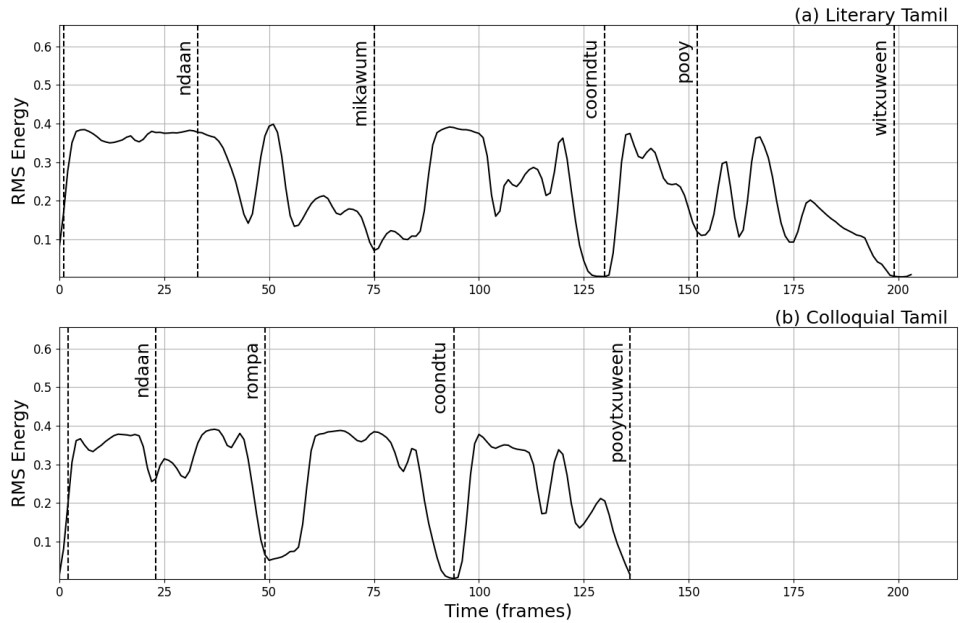

(b) Sentence 5

Figure 2: Energy contour of a pair of parallel LT and CT sentences

## 5.2 Voice Quality Features

Natural conversational speech consists of several voice qualities caused by non-modal phonations [41]. Voice quality features, and prosodic features, can be used to extract paralinguistic information such as speech acts, attitudes, and emotions. A perceptual and acoustic analysis on monosyllabic utterances spoken in different speaking styles, show that voice quality features are helpful in the identification of utterances with attitudinal and emotional expressions [42].

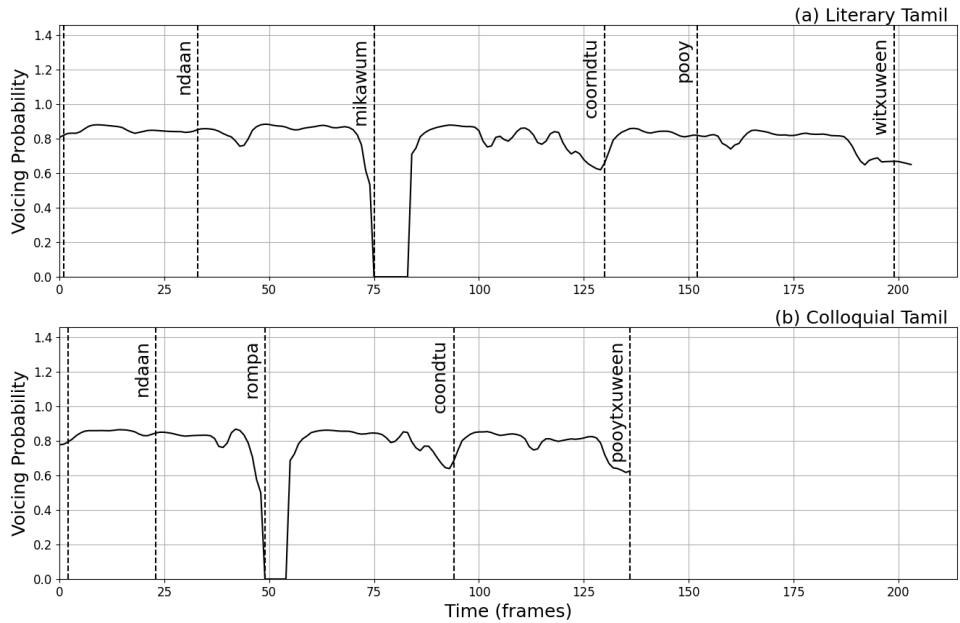

(a) Sentence 5

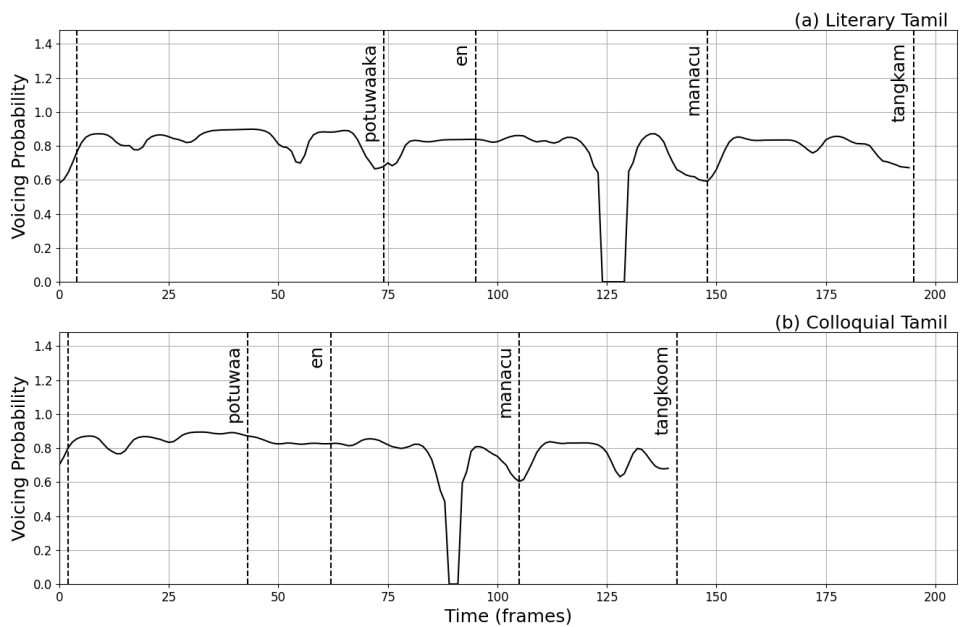

(b) Sentence 10

Figure 3: Voicing probability contour of a pair of parallel LT and CT sentences

The voice quality features considered in the current work include, jitter, derivative of jitter, shimmer, and harmonic-to-noise ratio. Some voice quality features can be thought of as micro-prosodic variations in speech. For example, jitter describes the variation of $F_0$, and shimmer describes the variation in $F_0$ amplitude. They are useful in describing vocal characteristics [43]. Jitter and shimmer can also be considered aperiodicity features, and are helpful in the detection of pathological speech [43, 44]. In a speaker verification task carried out using the Switchboard-I corpus, they are shown to be useful and complementary to spectral and prosodic parameters [45]. Similarly in the classification of human speaking styles, and animal vocalization and arousal levels, jitter and shimmer features are combined with spectral features and found to be useful [46] Harmonic-to-noise ratio can be thought of as another perspective to voicing probability. It is a measure of additive noise in the signal [47], and hence is able to determine the quality of vocal output. It is applied to tasks such as finding the degree of vocal aging [47], and laryngeal or voice pathologies [48].

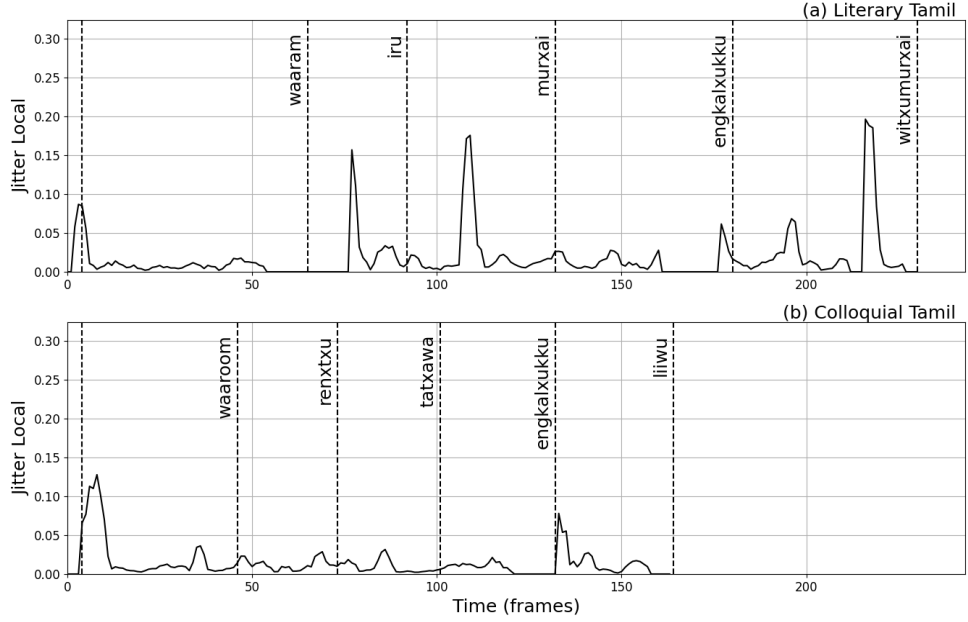

(a) Sentence 2

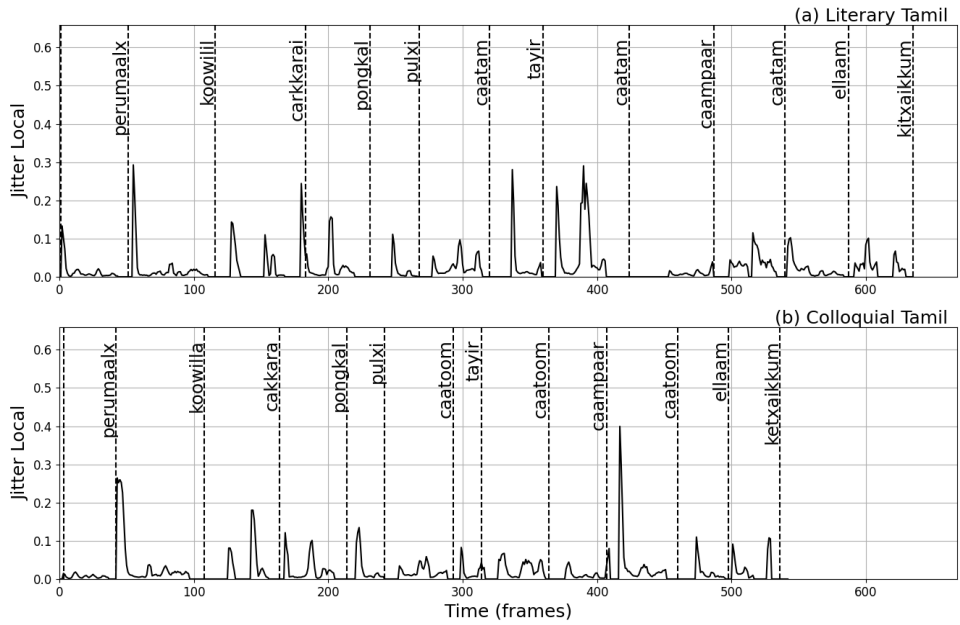

(b) Sentence 7

Figure 4: Jitter contour of a pair of parallel LT and CT sentences

### 5.2.1 Jitter

Jitter can be defined as the variation in the duration of two consecutive fundamental periods. The absolute local jitter can be given as,

$$J(n_p) = |T_0(n_p) - T_0(n_p - 1)| \quad for \quad 1 < n_p < N_p \tag{4}$$

where, $T_0$ denotes the pitch period, and $N_p$ is the total number of pitch periods estimated in a particular frame. The jitter for each frame is computed using a normalized (independent of underlying pitch period) and averaged jitter measure as,

$$\hat{J} = \frac{\frac{1}{N-1} \sum_{n_p=2}^{N_p} |T_0(n_p) - T_0(n_p - 1)|}{\frac{1}{N_p} \sum_{n_p=1}^{N_p} T_0(n_p)} \tag{5}$$

Figure 4 shows the jitter contour of a pair of parallel LT and CT sentences, which are sentences 2 and 7 in Tables 2 and 3 respectively. It can be observed from the Figures that LT is characterized by high rate of change of the fundamental frequency. This can be attributed to the expressive nature of literary Tamil, and the effortless speaking style of colloquial Tamil.

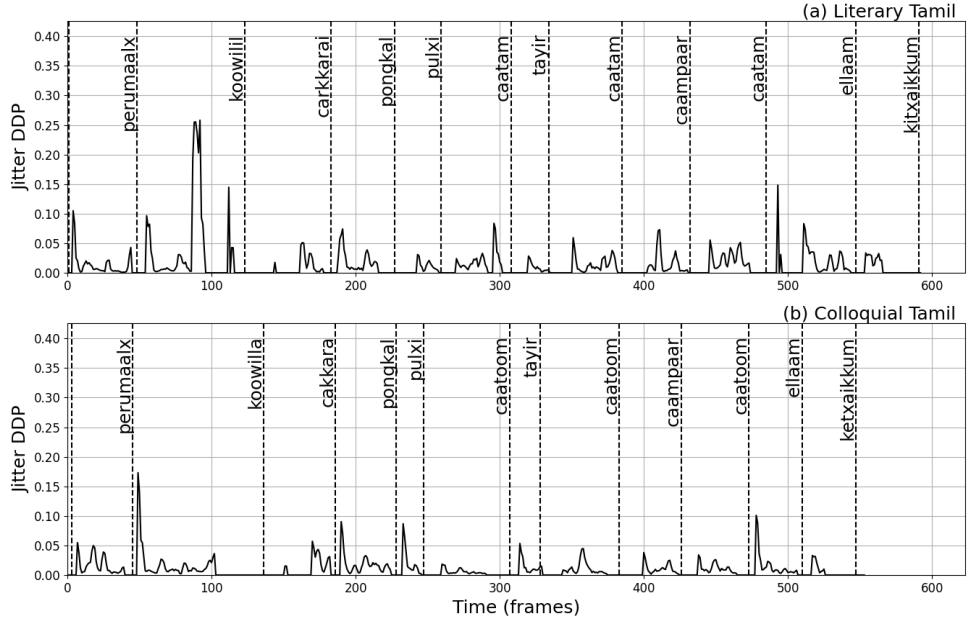

(a) Sentence 7

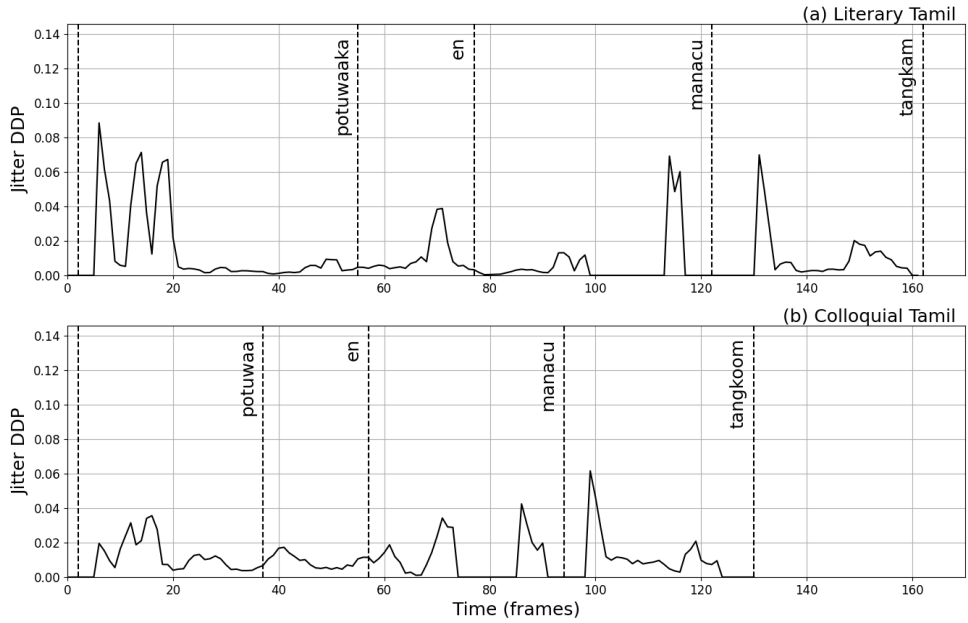

(b) Sentence 10

Figure 5: Derivative of Jitter contour of a pair of parallel LT and CT sentences

### 5.2.2 Derivative of Jitter

The 'derivative of jitter' is the variation of jitter across frames, or can be described as the 'jitter of jitter'. It is defined as,

$$J_d = |J(n_p) - J(n_p - 1)| \quad for \quad 2 < n_p < N_p. \tag{6}$$

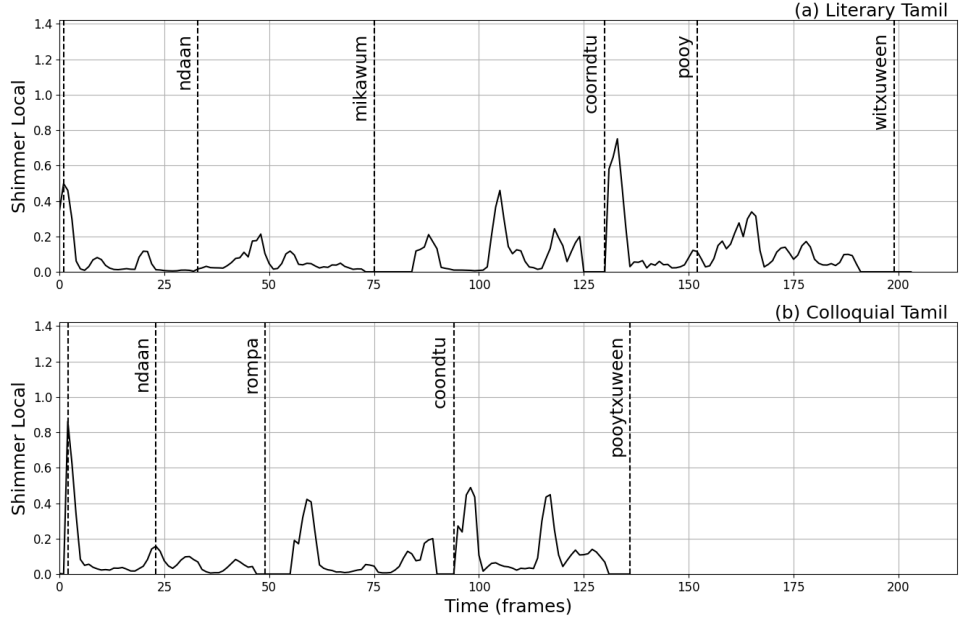

(a) Sentence 5

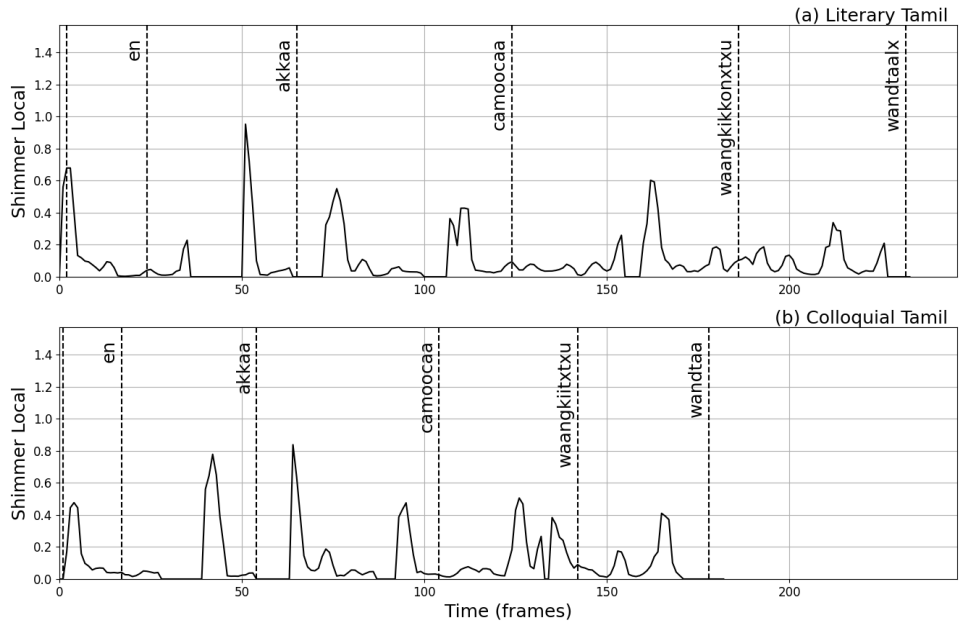

(b) Sentence 8

Figure 6: Shimmer contour of a pair of parallel LT and CT sentences

. The derivative of jitter can be computed for each frame using a normalized and averaged measure as,

$$\hat{J} = \frac{\frac{1}{N-2}\sum_{n_p=3}^{N_p}|J(n_p) - J(n_p-1)|}{\frac{1}{N_p}\sum_{n_p=1}^{N_p}T_0(n_p)} \quad (7)$$

where,

$$J(n_p) = |T_0(n_p) - T_0(n_p-1)|,$$
$$\text{and} \quad J(n_p-1) = |T_0(n_p-1) - T_0(n_p-2)|. \quad (8)$$

Figure 5 shows the 'derivative of jitter' contour of a pair of parallel LT and CT sentences, which are sentences 7 and 10 in Table 3. In Figure 4, the jitter of sentence 7 is plotted, which shows the rate of change of fundamental frequency. While in Figure 5, the rate of change of jitter is shown. It can be seen from Figures 4 and 5 that the derivative of jitter conveys more information than jitter.

### 5.2.3 Shimmer

Shimmer as a feature is useful where ever jitter is used [42, 43]. It describes the amplitude variations of consecutive speech signal periods. It is given as,

$$S_p = |A(n_p) - A(n_p - 1)|, \tag{9}$$

where, $A$ is the peak to peak amplitude within a pitch period.

Figure 6 shows the shimmer contour of a pair of parallel LT and CT sentences, which are sentences 5 and 8 in Tables 2 and 3 respectively. From the Figure it can be seen that LT is characterized by a large number of minor peaks. Similar to how the fundamental frequency has varied for jitter in Figure 4, in Figure 6, there is variation in amplitude, for each fundamental period. Similar to how quick variations of the fundamental frequency in LT is manifested in the jitter contour plots, the variations in amplitude for each fundamental period, is manifested in the shimmer plot.

### 5.2.4 Harmonic-to-Noise Ratio

Harmonic-to-Noise ratio is useful for capturing the hoarseness in speech [49], and is similar to the measure of voicing probability. It is defined as the ratio of the energy in the harmonic signal components ($E_h$), to the energy in the noise-like signal components ($E_n$), given by,

$$HNR = \frac{E_h}{E_n}. \tag{10}$$

Based on this definition it can be seen that HNR is a measure of efficiency of speech production, that is, it measures the proportion of energy expelled from the lungs that is converted to pitch and pitch harmonics. HNR is also related to voicing probability, in that both of them quantify voiced-ness in speech. In this way, it relates the physiological aspects of speech production to it's perception.

Figure 7 shows the log-HNR contour of a pair of parallel LT and CT sentences, which are sentences 9 and 10 in Table 3. Comparing voicing probability and HNR of sentence 10 in Figure 3 and 7 respectively, it can be observed that HNR is more sensitive to unvoiced sounds than voicing probability which drops in value only for the fricative /s/. The familiar notions of CT contours being smoother than LT can also be observed in Figure 7.

## 5.3 Spectral Features

Spectral features describe the spectral characteristics of the signal, and are derived from the (discrete Fourier transform derived) magnitude spectrum. The chosen spectral features address both static and dynamic spectral characteristics. Since loudness and pitch related characteristics are addressed by prosodic and voice quality features, spectral features can be useful in describing the timbre of an utterance [6]. Two spectral features are considered in the current work. They are, spectral flux, and psychoacoustic sharpness. Cepstral features are not included in this set.

### 5.3.1 Psychoacoustic Sharpness

Psychoacoustic sharpness is a static descriptor derived from a single frame and is based on the spectral centroid. The spectral centroid is given as,

$$S_c = \frac{\sum_{k=1}^{K/2} F(k) X(k)}{\sum_{k=1}^{K/2} X(k)} \tag{11}$$

where, $X(k)$ is the magnitude spectrum, $F(k)$ is the frequency in Hz corresponding to bin $k$, and $K$ is the size of the FFT. The spectral centroid becomes a measure of the "perceived" sharpness of a sound, when it is computed on a Bark scale, and is given as,

$$S_s = \frac{\sum_{k=1}^{K/2} \Theta_b(F(k)) X(k)}{\sum_{k=1}^{K/2} X(k)} \tag{12}$$

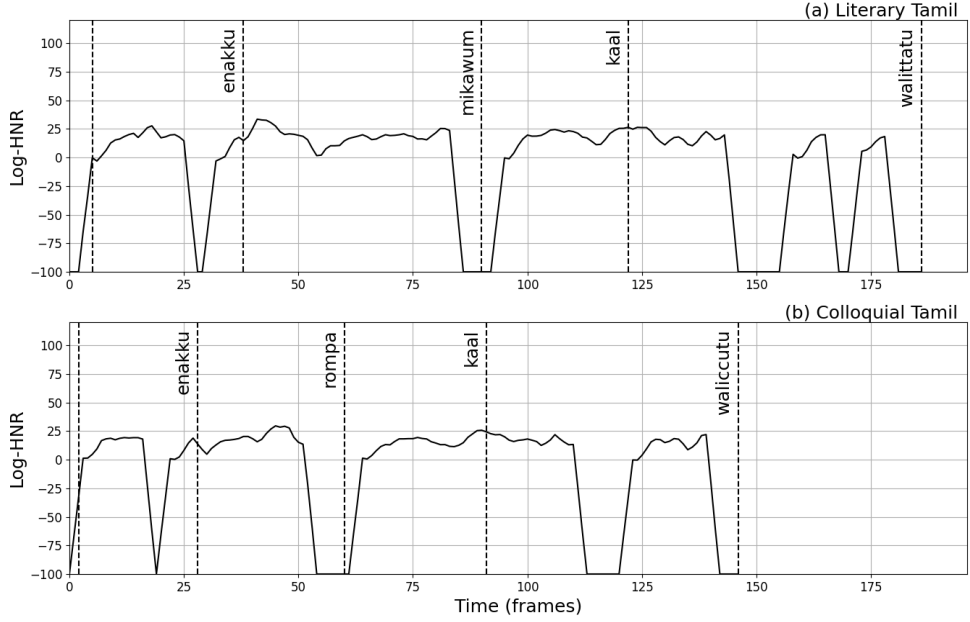

(a) Sentence 9

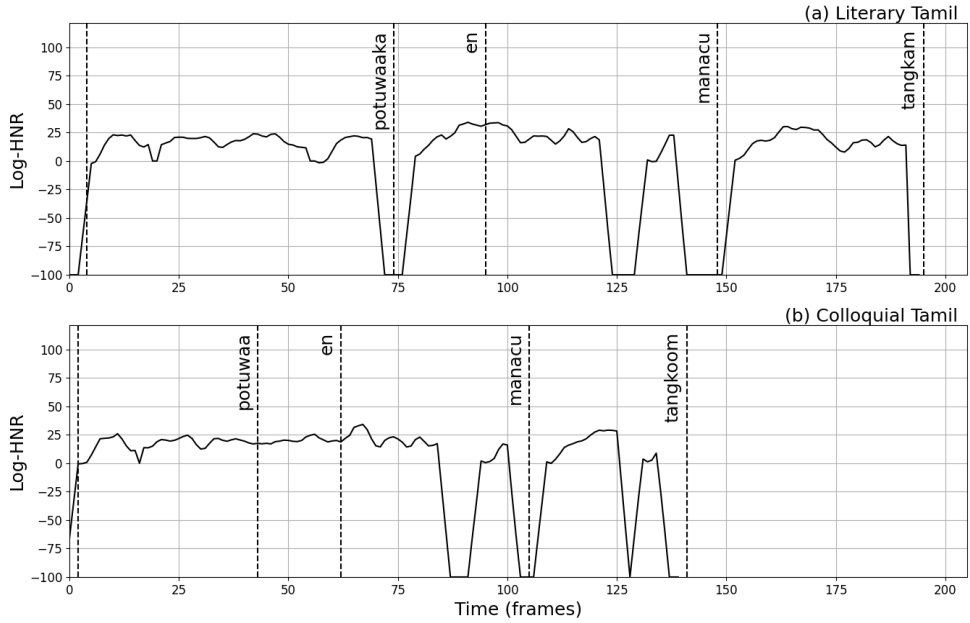

(b) Sentence 10

Figure 7: Contour of the log-Harmonic-to-Noise ratio, for a pair of parallel LT and CT sentences.

where $\Theta_b$ is a forward transformation function that converts linear frequency scale to the Bark scale.

Figure 8 shows the contour of psychoacoustic sharpness for a pair of parallel LT and CT sentences, which are sentences 5 and 10 in Tables 2 and 3 respectively. Comparing the voicing probability and psychoacoustic sharpness of sentence 10 in Figure 3 and 8 respectively, it can be observed that they are fairly opposite to each other. This can be attributed to the fact that the sharpness in sound is mainly contributed by high frequencies which are present in unvoiced fricatives such as /s/ in the word 'sorndhu (சோர்ந்து)' in sentence 5, and the word 'manasu (மனசு)' in sentence 10. Hence, peaks can be found in unvoiced segments of the contour of psychoacoustic sharpness. Finally, similar to the other cases, the contour of CT is smoother than that of LT.

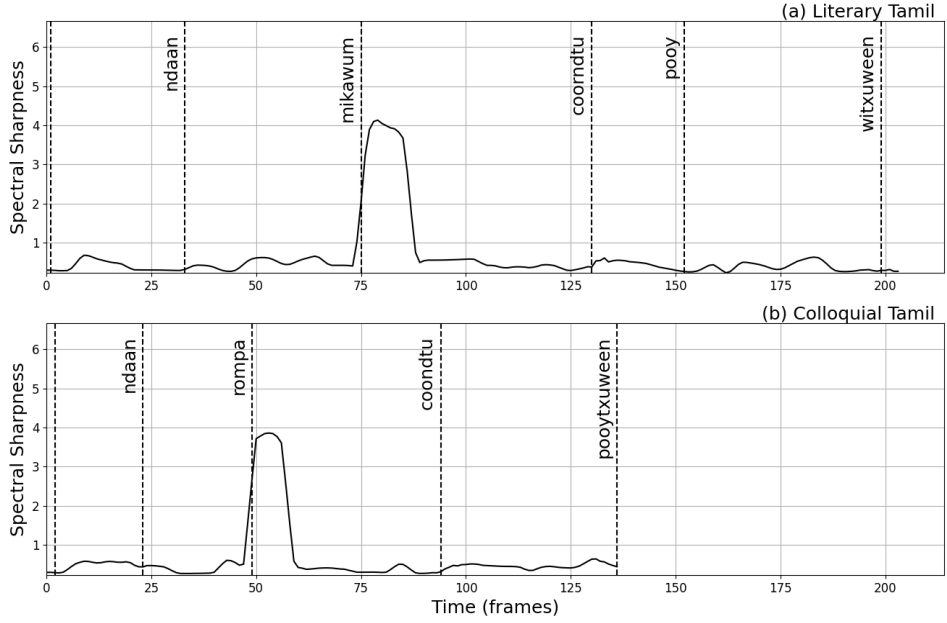

(a) Sentence 5

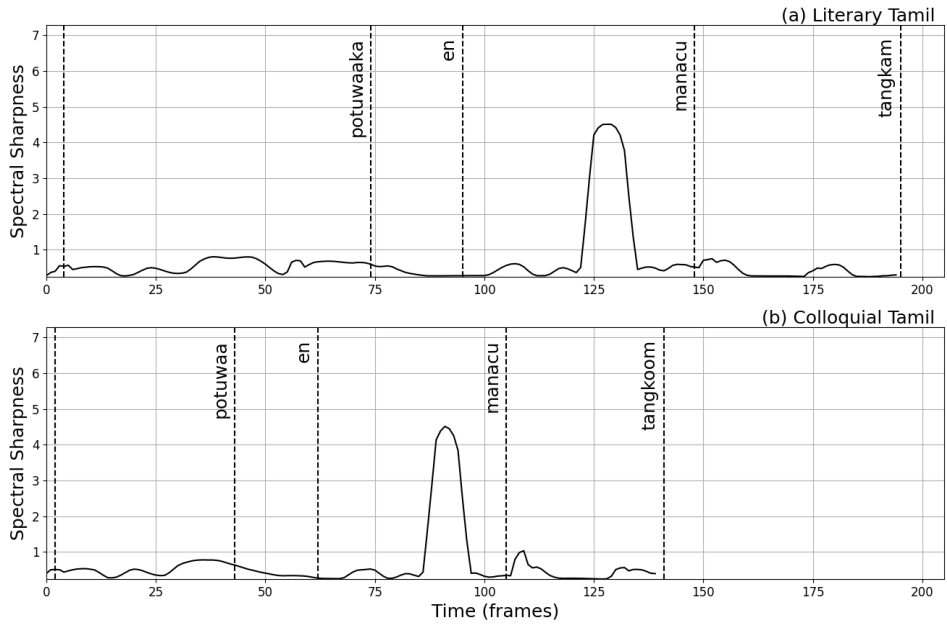

(b) Sentence 10

Figure 8: Contour of the psychoacoustic sharpness, for a pair of parallel LT and CT sentences.

### 5.3.2 Spectral Flux

Spectral flux is a dynamic feature, in contrast to features such as spectral centroid, spectral slope, and spectral entropy, which are static. It addresses the change in the spectrum over time. It is an useful measure to determine whether the spectral characteristics of a signal vary slowly or quickly over time.

Spectral flux is defined as the normalized spectral difference between two successive frames, and is given as,

$$S_f(m) = \sum_{k=1}^{K/2} \left( \frac{X(m,k)}{\mu_m} - \frac{X(m-1,k)}{\mu_{m-1}} \right)^2 \qquad (13)$$

where, $X(m,k)$ is the short-time spectrum, $K$ the size of the FFT, $\mu_m$ and $\mu_{m-1}$ the normalization coefficients for the $m^{th}$ and $(m-1)^{st}$ frames respectively. The normalization coefficient $\mu_m$ is given as,

$$\mu_m = \sqrt{\sum_{k=1}^{K/2} X(m,k)^2} \tag{14}$$

Figure 9 shows the spectral flux contour of a pair of parallel LT and CT sentences, which are sentences 3 and 5 in Table 2. The degree of change in spectral characteristics of the signal can be observed in Figure 9 for sentence 5, where quick changes in the feature contour can be observed for LT. This can be attributed to the fact that for a given time period, LT requires more change in the vocal tract configurations, when compared to CT. Another aspect that can be observed from the contour corresponding to sentence 3 is that of the nasalized vowel of the word 'bayo(m) (பயோ(ம்))', which manifests as a flattened contour in CT.

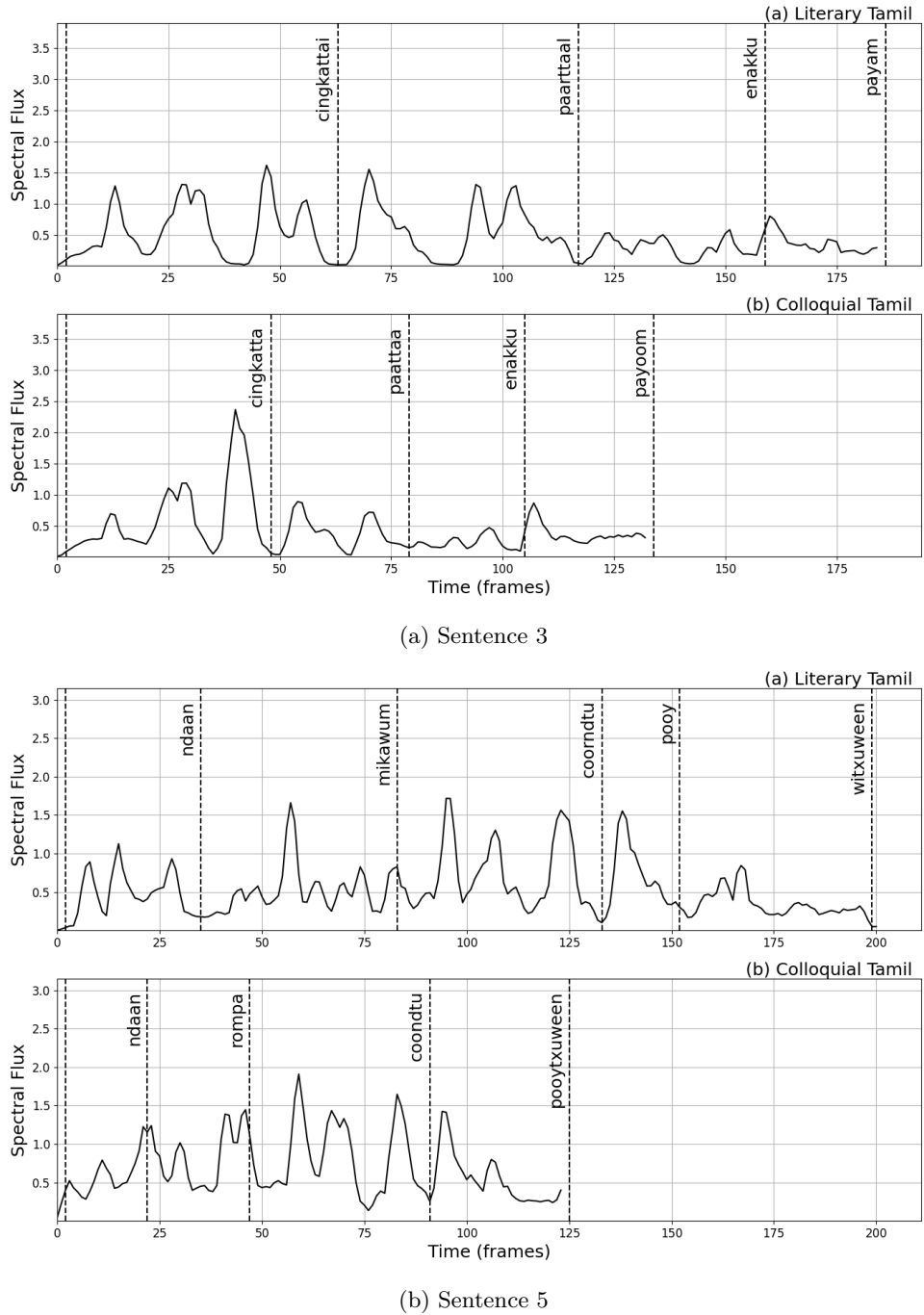

(a) Sentence 3

(b) Sentence 5

Figure 9: Contour of the spectral flux, for a pair of parallel LT and CT sentences.

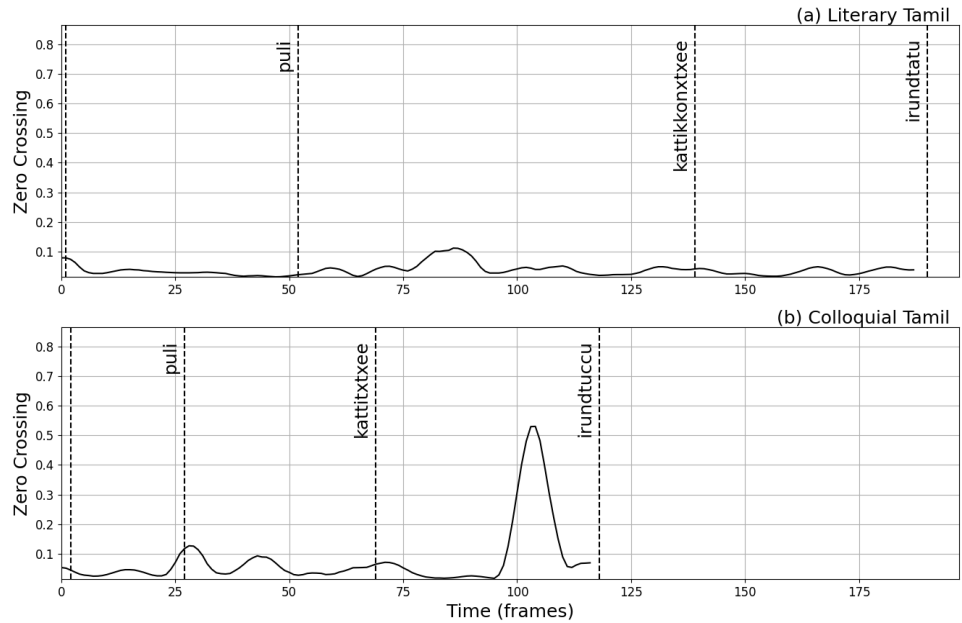

(a) Sentence 4

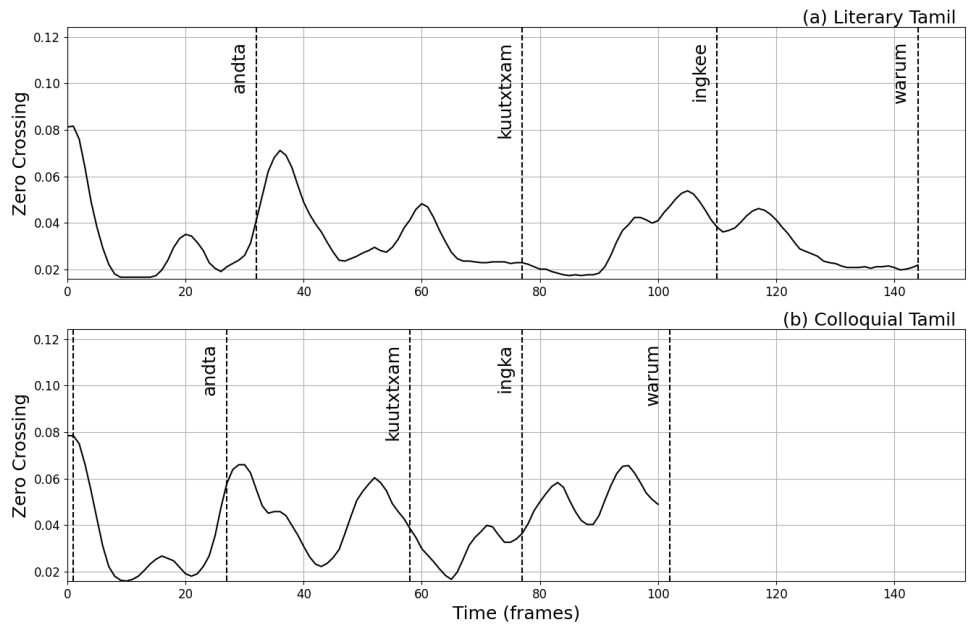

(b) Sentence 6

Figure 10: Contour of the zero crossing rate, for a pair of parallel LT and CT sentences.

## 5.4 Temporal Feature

Temporal features provide meaningful information about the temporal patterns, trends, or dynamics present from time-dependent data, allowing for better understanding, interpretation, and prediction of temporal phenomena.

### 5.4.1 Zero Crossing Rate

Zero crossing rate (ZCR) is a versatile feature, used in voiced-unvoiced classification [50], speech-music classification, voice activity detection [51], emotion recognition [52], and in the analysis, segmentation and recognition of speech sounds. Zero-crossing rate is defined as the number of sign changes in a signal $x(n)$, per unit of time, or the rate of sign change. It is expressed in unit of crossings per second. A crossing $c$ is defined as,

$$c = \begin{cases} x(n-1)x(n) < 0 & x(n) \neq 0 \\ x(n-1)x(n+1) < 0 & x(n) = 0 \end{cases} \quad (15)$$

Figure 10 shows the contour of the zero crossing rate for a pair of parallel LT and CT sentences, which are sentences 4 and 6 in Tables 2 and 3 respectively.

## 5.5 Conclusions on Handcrafted Features

A common trend that can be observed in the contours of the handcrafted features corresponding to literary Tamil speech is the higher level of granularity in the form of quick changes in the contour, while for CT, the opposite is true. It can further be noticed that the duration of CT in all the plots is lower than that of LT. This is due to the fact that CT avoids intricate details present in the LT (refer Section 4) by,

- avoiding effortful phonemes,
- avoiding consonant at the end of the word, and attaching to the next word,
- avoiding nasal consonants and converting the previous vowel into a nasalized vowel, and
- avoiding unvoiced stops.

Although the proposed 1DCNN does not learn duration as a feature, these characteristics of CT produces connected speech, which causes the smooth and shrunk contour of the handcrafted features, which is learnt by the 1DCNN.

## 5.6 MFCC

MFCC is one of the most popular and ubiquitous feature, that is vastly utilized in various applications such as, speech recognition, speaker recognition, emotion recognition, language and dialect identification, speech classification, music information retrieval, music genre classification, and music instrument identification. The sequence of vocal tract shapes that correspond to a particular speech utterance, contain dialect-relevant pronunciation cues. MFCC features capture these vocal tract configurations. Learning the trend of these MFCC features is analogous to learning the trend in vocal tract configurations.

# 6 Proposed Approach for LCTID

The proposed 1DCNN architecture is trained on the features detailed in the previous section. These features are extracted from the utterances of the Microsoft speech corpus, which was featured in [53]. The proposed 1DCNN architecture, and the way in which the Microsoft speech corpus is adapted for evaluating literary and colloquial dialect identification is detailed in the sections that follow.

## 6.1 CNN Architecture

As discussed in Section. 3.2, the 1D-CNN has many distinct advantages for time series data. All CNNs, whether 1D or 2D, have feature extraction capabilities. The features being learnt by the CNN can also be considered as intermediate features, if external features are the input to the network. Providing external features makes the system more interpretable and controllable, whilst also converging faster (when compared to training with raw signals). In the proposed approach, handcrafted and MFCC features are extracted from the speech signal and are fed as input to the 1D-CNN.

The role of the 1D-CNN here is to learn the trend of the proposed features over time. A detailed description of the forward- and back-propagation in 1D-CNN can be found in [54], a summary is provided below. The input $x_k^l$, to a node $k$, in convolutional layer $l$, is given by,

$$x_k^l = \sum_{i=1}^{N_{l-1}} w_{ik}^{l-1} * s_i^{l-1} + b_k^l \quad (16)$$

where, $b_k^l$ is the bias of node $k$ in layer $l$, $w_{ik}^{l-1}$ is a set of weights connecting the nodes of the previous layer $l-1$ with the node $k$ in the current layer $l$, $s_i^{l-1}$ is a set of outputs from the previous layer $l-1$, and $*$ denotes the one-dimensional convolution operation. The number of input features is greater than one, hence they are considered as channels of the input to the 1D-CNN, and the convolution operation is still performed over a single dimension (that is, across time). The filters are learnt across the multiple channels, and then combined.

Table 4: Various architectures of the 1DCNN.

|  | **CA01** | **CA02** | **CA03** |
|---|---|---|---|
| Convolutional Layers | Two sets of two convolutional layers (4 layers) | Two sets of two convolutional layers (4 layers) | Two sets of two convolutional layers (4 layers) |
| Kernel Sizes | [10, 10, 5, 5] | [7, 7, 3, 3] | [7, 7, 3, 3] |
| Dense Layers | [1024] | [1024] | [1024, 512] |
| Training Method | Mini-Batch Gradient Descent | Mini-Batch Gradient Descent | Stochastic Gradient Descent |

The generic architecture of the proposed 1D-CNN is illustrated in Figure 11. This architecture consists of two sets of two convolutional layers each, followed by dense layer(s). Each set of convolutional layer is followed by a maxpooling and dropout layer. All layers are activated by a rectified linear unit (ReLU) activation function, except the last layer which is activated by a soft-max activation function. Three variations of this architecture (CA01, CA02, and CA03), that are proposed for evaluation, are presented in Table 4.

```
   OPERATION            DATA DIMENSIONS   WEIGHTS(N)   WEIGHTS(%)

       Input     #####       187    10
      Conv1D      \|/  -------------------      2272        0.1%
        relu     #####       181    32
      Conv1D      \|/  -------------------      7200        0.3%
        relu     #####       175    32
 MaxPooling1D    Y max -------------------         0        0.0%
                 #####        87    32
     Dropout      | || -------------------         0        0.0%
                 #####        87    32
      Conv1D      \|/  -------------------      6208        0.2%
        relu     #####        85    64
      Conv1D      \|/  -------------------     12352        0.5%
        relu     #####        83    64
 MaxPooling1D    Y max -------------------         0        0.0%
                 #####        41    64
     Dropout      | || -------------------         0        0.0%
                 #####        41    64
     Flatten     ||||| -------------------         0        0.0%
                 #####      2624
       Dense     XXXXX -------------------   2688000       98.9%
        relu     #####      1024
     Dropout      | || -------------------         0        0.0%
                 #####      1024
       Dense     XXXXX -------------------      2050        0.1%
     softmax     #####         2
```

Figure 11: Proposed 1D-CNN architecture

It is conventionally known that if meaningful prosodic features are to be extracted, they have to be supra-segmental. Hence, it is useful to consider at least the duration of a syllable when extracting prosodic information. This gives us a preamble to meaningfully set the size of the 1D-CNN kernels. In [11], authors propose an LID system that is trained on short-term and long-term components (segments). The long-term component captures the prosodic movements over several psuedo-syllables, while the short-term component captures the prosodic movements inside a psuedo-syllable, where a psuedo-syllable is the segment between one vowel onset point and another. Here, it is shown that the short-term components provide better results than the long-term components. These arguments provide a foundation for the size of the kernel that is to be used in the 1DCNN. The kernel sizes used in the proposed 1DCNN architecture are 10 and 5 in the first and second set of convolutional layers respectively. These kernel sizes amount to a duration of 150 and 100ms respectively when the framesize, for extracting the features, is set to 60ms for prosodic features, and a duration of 110 and 60ms respectively when the framesize is set to 20ms for all other features. Hence, in the current work, the classification framework handles long-range temporal-dependencies, and the features also summarize the information over a meaningful unit of time.

## 6.2 Training Methodology and Inference

CNNs require fixed sized inputs. However, naturally occurring time series data almost always have variable lengths, and the same is true with speech utterances. Figure 12 shows the probability density function of the utterance durations in the Microsoft speech corpus, which illustrates this variation. In these cases, all the training examples must be normalized to the same duration (normalized duration), and fixing the duration of the utterance requires some thought. Several concerns arise:

- If the normalized duration is set based on the utterance with the maximum duration, then it would mean that almost all utterances would have to be zero padded. Excessive zero padding may bias the system into learning the patterns of zeros, rather than the features themselves.

- If this normalized duration is set based on the utterance with the minimum duration, then it would mean that almost all utterances will be truncated. This means that we voluntarily disregard large amounts of information that can be used for classification.

- Yet another case would be to balance these two cases and empirically determine the normalized duration such that valuable information is not lost, while at the same time excessive zero padding is controlled.

In the current work, we take another approach and propose that each utterance be split into multiple segments, based on a pre-determined segment duration ($d_s$). This duration is established to be the first (lower) quartile of the normally distributed duration of all utterances in the training set, which amounts to 1.87s. The first quartile is defined as the $((n+1)/4)^{th}$ term, in a set of $n$ observations (here, durations) that are arranged in an ascending order. With the segment duration established, a single training utterance is first split into multiple segments of duration $d_s$, with the last segment being zero padded as required to match the segment length. The number of segments $n_s$ associated with each utterance is given by, $n_s = \lceil d_u/d_s \rceil$, where $d_u$ is the duration of the utterance. The duration of zero padding $d_z$ can be given as, $d_z = d_s - d_l$, where $d_l$ is the duration of the last segment.

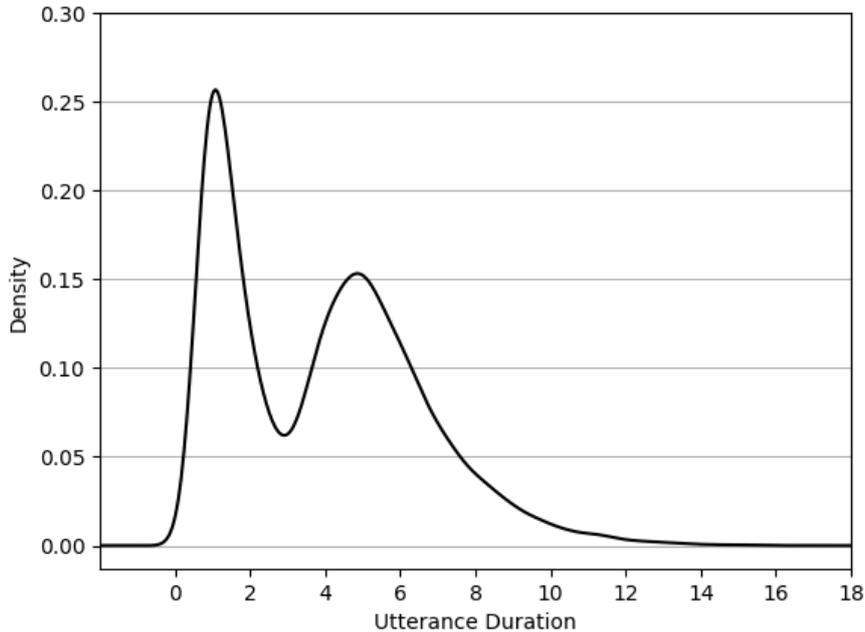

Figure 12: The probability density function of the utterance durations in the Microsoft speech corpus.

During model training, each of these segments are assigned the label of the utterance from which these segments were derived. During inference, the dialect of a test utterance is determined by first averaging the activations corresponding to each dialect, across all segments, and then selecting the dialect which has the higher average. Hence, the predicted dialect $\hat{D}$ can be given by,

$$\hat{D} = arg \max_{D=lt,ct} \sum_{s=1}^{n_s} \sigma_D \tag{17}$$

where, $\sigma_D$ corresponds to the activation of the 1DCNN model corresponding to dialect $D$ in segment $s$.

### 6.3 Dataset

The dataset proposed for evaluation in the current work is the Microsoft speech corpus, in contrast to our previous works [5, 55], where the smaller in-house corpus was used. The Microsoft speech corpus was originally created for the purpose of building automatic speech recognition systems, and as such contains a large amount of speech data. Upon analysis, it was found that the dataset contains a collection of both read and spontaneous speech utterances, most probably for inclusivity. The read speech in the dataset are monologues which are recorded in a studio environment. This read speech can be considered analogous to literary Tamil, because in contemporary reality, literary Tamil is mostly spoken in a formal setup, as a monologue. The spontaneous speech in the dataset are monologues and dialogues which are recorded in a live environment. This spontaneous speech can be considered analogous to colloquial Tamil, because in contemporary reality, colloquial Tamil is mostly spoken in an informal setup and between two or more people. Even though both read and spontaneous speech were present in the dataset, there were no labels that could be readily used for evaluation. Hence by manually sorting through the dataset, and determining the labels, this dataset has been repurposed for the classification of LT and CT. The details of the dataset post manual sorting are provided in Table 5.

Table 5: Details of the Microsoft Dataset

|  | **LT** | **CT** | **Total** |
|---|---|---|---|
| **No. of Utterances** | 19612 | 19519 | 39131 |
| **Duration (Hours)** | 31.49 | 8.11 | 40 |

## 7 Experimental Results and Observations

The evaluation of the proposed approach is detailed in this section. First, the baseline results of the proposed 1DCNN architectures trained with the proposed handcrafted and MFCC features, extracted from the Microsoft speech corpus, are discussed. Following this, the proposed feature ablation and the feature combination studies are discussed.

### 7.1 Baseline Experiment

In this baseline experiment, the proposed 1DCNN architecture is trained with two sets of features: (i) the ten chosen handcrafted features, and (ii) the ubiquitous MFCC features. Three variations of the proposed 1DCNN architecture, as shown in Table 4, are evaluated using these two sets of features.

The results of this experiment are detailed in Table 6. In each case, the Precision (Pre), Recall (Re), F1 Score (F1) and the over all accuracy (Acc), are reported. It can be seen that CA02 is better than CA01. This can be atributed to the size of the kernels used in CA02. This kernel size is retained in CA03, which performs the best. This can be attributed to two reasons,

1. The number of fully connected layers has been increased to two, instead of one. This enables the architecture to learn a non-linear hyperplane.

2. Mini-batch gradient descent is replaced with stochastic gradient descent, wherein the samples in each batch are picked randomly, enabling the loss function to navigate to the global minima steadily.

The dataset used for training CA01, 02 and 03, is imbalanced, where the total duration of LT (31 hours) is more than CT (8 hours), as shown in Table 5. Hence, a balanced dataset is derived, with only 8 hours in both LT and CT. CA03 trained on this balanced dataset is named CA03B in Table 6. It can be observed from the Table that the F1 scores (for both the classes), and the over all accuracy improves when CA03 is trained on this balanced dataset. This is the case for both feature sets. More specifically, it can be seen that the recall of CT has improved, meaning more utterances have been identified as CT by the system. This can be directly attributed to the fact that the dataset is balanced, hence reducing the bias towards LT in the previous three systems. Another point to notice here is that even though the duration of the dataset is reduced drastically, the performance of the system was not affected. Henceforth, in all experiments that follow, CA03 architecture trained on balanced data (that is, CA03B) is used.

Table 6: Results of Baseline Experiments using the three variations of the proposed architecture.

| Arch | Feature Set | LT | | | CT | | | Overall |
|---|---|---|---|---|---|---|---|---|
| | | Pre | Re | F1 | Pre | Re | F1 | Acc |
| CA01 | Handcrafted | 0.9580 | 0.9930 | 0.9752 | 0.9913 | 0.9477 | 0.9690 | 0.9724 |
| | MFCC | 0.9670 | 0.9986 | 0.9826 | 0.9983 | 0.9591 | 0.9783 | 0.9807 |
| CA02 | Handcrafted | 0.9701 | 0.9863 | 0.9781 | 0.9832 | 0.9636 | 0.9733 | 0.9760 |
| | MFCC | 0.9713 | 0.9883 | 0.9797 | 0.9857 | 0.965 | 0.9752 | 0.9777 |
| CA03 | Handcrafted | 0.9623 | 0.9915 | 0.9767 | 0.9894 | 0.9534 | 0.9711 | 0.9742 |
| | MFCC | 0.9709 | 0.9997 | 0.9850 | 0.9996 | 0.9640 | 0.9815 | 0.9835 |
| CA03B | Handcrafted | 0.9824 | 0.9815 | **0.9819** | 0.9779 | 0.9788 | **0.9784** | **0.9803** |
| | MFCC | 0.9828 | 0.9981 | **0.9904** | 0.9977 | 0.9791 | **0.9883** | **0.9895** |

## 7.2 Feature Ablation

Feature ablation assists in understanding the importance of features by removing those that do not contribute to the model. To carry out the feature ablation study for the current set of ten handcrafted features, initially, recursive feature elimination (RFE) is used.

### 7.2.1 Recursive Feature Elimination (RFE)

RFE is a recursive algorithm in which, starting from the whole set of $n_f$ features, each round of evaluation $1 \leq r < n_f$ is carried out using $n_f - r + 1$ possible subsets of $n_f - r$ features, determining and eliminating the least performing feature in each round $r$. The number of rounds of evaluation can be reduced, if a certain number of desired features $n_d < n_f$ are required. The CA03B architecture is used, and in each round, four-fold cross validation is carried out.

Table 7: Outcomes in terms of Accuracy and Rank of each feature after one round of RFE.

| Eliminated Feature | Accuracy | Rank |
|---|---|---|
| Fundamental Frequency | 0.9907 | 4 |
| Voicing Probability | 0.9913 | 8 |
| Energy | 0.9908 | 5 |
| Zero Crossing Rate | 0.9916 | 9 |
| Harmonic-to-Noise Ratio | 0.9901 | 2 |
| Derivative of Jitter | 0.9896 | 1 |
| Jitter | 0.9909 | 7 |
| Shimmer | 0.9908 | 5 |
| Spectral Flux | 0.9906 | 3 |
| Psychoacoustic Sharpness | 0.9916 | 9 |

Table 7 shows the outcome of the first round of RFE evaluation in terms of accuracy and rank of each feature when each feature is eliminated in the first round of evaluation. Here, lower accuracy implies more impact of the feature in the classifier, and hence a better rank. From Table 7 the following observations are made in terms of the accuracies:

- There is not much variation in the values, when each of the ten handcrafted features are ablated. For example, consider the values when the following features are ablated: energy (acc: 0.9908, rank: 5), jitter (acc: 0.9909, rank: 7) and shimmer (acc: 0.9908, rank: 5).
- The values are sometimes equal, as in the previous example.
- The values vary over a small range from 0.9896 to 0.9916.

These observations lead to the conclusion that all these features contribute almost equally to the results. Figure 13 shows a radial plot with the accuracies offered by the system when each feature is ablated. It can be seen from the Figure that, due to the narrow range of values in the y-axis from 0.9 to 1, it forms a near perfect circle, indicating how close the values are to each other. Furthermore, it can be noticed that two features have the same last rank, and the difference between the last two ranks (9 and 8) is 0.03%, making it impossible to reliably determine the least performing feature. Selecting one randomly would, by principle, go against the method of recursive feature elimination. These observations, although speaking positively of the relevance of the proposed hand-crafted features, do not provide conclusive evidence to ablate the least performing feature, and to move forward towards the second round of RFE. To deal with this, a straight-forward and counter-intuitive approach (to RFE), which works particularly well for the current task and the hand-crafted features, is proposed in the next section.

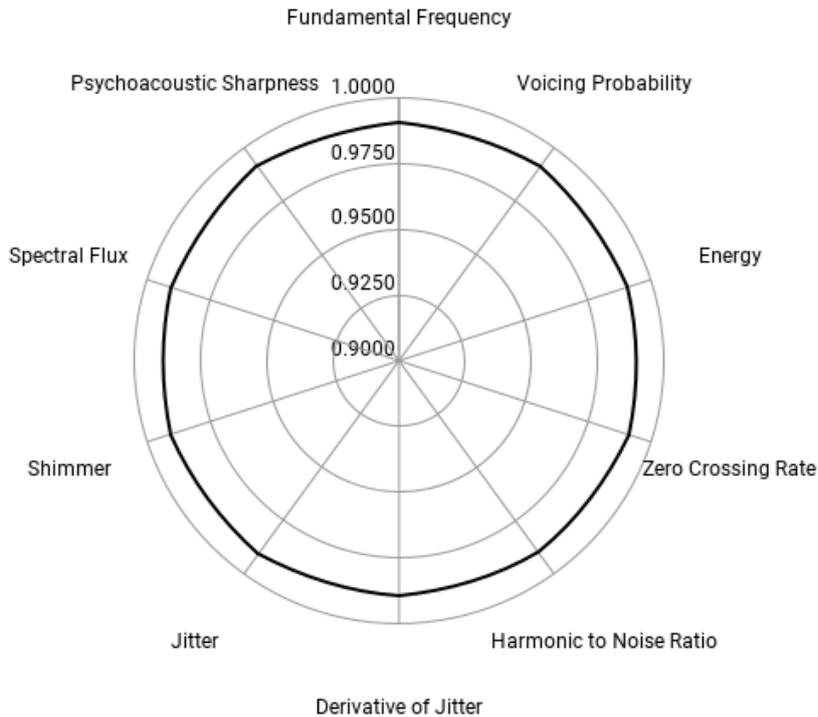

Figure 13: Radial plot showing the accuracies obtained in round one of RFE, when each specified feature is ablated. Since the range of the values is narrow, they form a near perfect circle, despite the narrow range of the y-axis.

### 7.2.2 Independent Feature Evaluation (IFE)

In IFE, the importance of each feature is determined by doing the opposite of the RFE method, that is by evaluating only one feature instead of $n_f - r$ features in each round of evaluation. The advantage of this method is that, it requires only $n_f$ number of evaluations, where as in RFE $n_f!$ number of evaluations are required, where $n_f$ is the total number of initial features. In RFE, the focus is on finding the least contributing feature, while in IFE the focus is on finding the most contributing feature (independent of other features). Table 8 presents the over all accuracy, and the ranks determined by the IFE method. The following observations can be made from this Table:

- The accuracies now fall in the range 0.9310 to 0.9922, a difference of 6.12%. A range that is much larger than RFE, from 0.9896 to 0.9916, with a difference of 0.20%.

- The accuracies are not as close to each other as in RFE. Table 9 shows the minimum, maximum, and mean differences in accuracy corresponding to features with subsequent ranks. It can be seen from the Table that the maximum difference in RFE is less than the minimum difference in IFE. This can be further observed when comparing Figures 13 and 14.

- There are no features with the same rank, hence providing a deterministic way to ablate features as required.

- The ranks in IFE offer a more conclusive evidence of the feature's contribution to the classification task, making it easier to determine the best and least effective features.

The conclusions of these experiments is that, the three best effective features are: energy, harmonic-to-noise ratio, and fundamental frequency, and the three least effective features are: jitter, zero crossing rate and psychoacoustic sharpness. These features are further pursued in feature combination experiments that follow.

Table 8: Performance and Rank of IFE System

| Feature | Accuracy | Rank |
| --- | --- | --- |
| **Fundamental Frequency** | 0.9906 | 3 |
| **Voicing Probability** | 0.9888 | 4 |
| **Energy** | 0.9922 | 1 |
| **Zero Crossing Rate** | 0.9582 | 9 |
| **Harmonic-to-Noise Ratio** | 0.9916 | 2 |
| **Derivative of Jitter** | 0.9837 | 5 |
| **Jitter** | 0.9777 | 8 |
| **Shimmer** | 0.9801 | 7 |
| **Spectral Flux** | 0.9825 | 6 |
| **Psychoacoustic Sharpness** | 0.9310 | 10 |

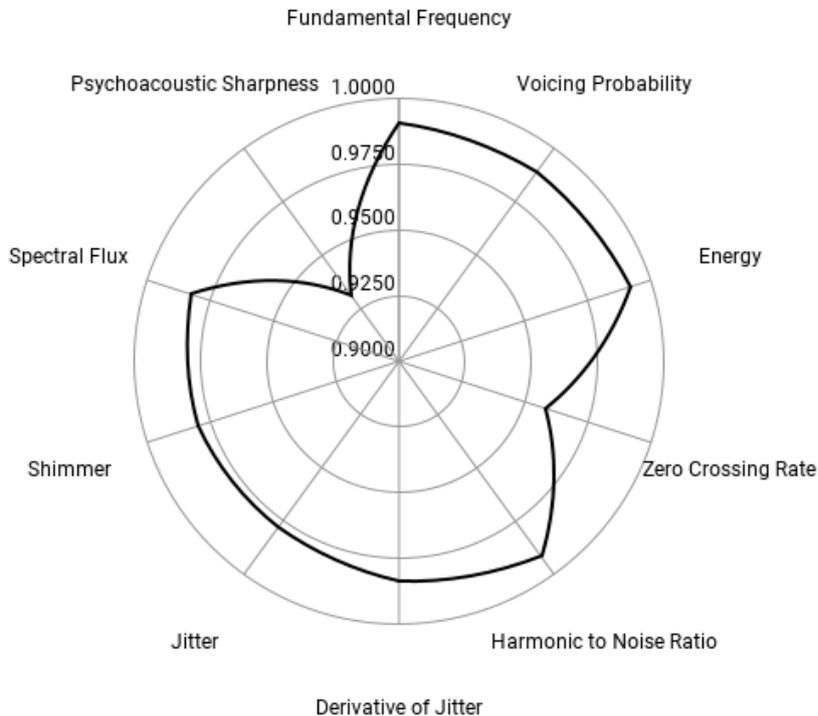

Figure 14: Radial plot showing the accuracies obtained from IFE, when each specified feature is considered.

## 7.3 Feature Combination

In the previous sections, first, baseline systems using the proposed 1D-CNN architecture trained on the proposed set of ten hand-crafted features, and also the MFCC, were evaluated. From this experiment it can be observed that MFCC features offer 0.92% improvement over the proposed set of hand-crafted features. These results naturally lead to the conclusion that MFCC may be enough. Yet, further investigation through the ablation study revealed that using the best features individually offered more results (example, 0.9922 for energy) than all features combined (0.9803), or the MFCC (0.9895). Furthermore, prosodic features in literature consistently offer inferior results when compared to MFCC, which makes the results obtained in the current work through handcrafted features more significant. For example, in [14] duration, pitch, and energy features offer accuracies of 64, 67 and 53%, while MFCC offers 73%, for dialect and emotion identification. In [16], pitch flux offers accuracies below 70%, while MFCC offers 90%, for a Chinese dialect identification task. In [39], it is shown that MFCC features offer better accuracies than prosodic features for language identification across three durations of the test utterances (3s, 10s, and 30s), and prosodic features extracted from three levels (syllable, di-syllable, word, and whole utterance). Better performance using handcrafted prosodic features in the current work can be attributed to the fact that the proposed 1DCNN learns the trend of the handcrafted features, which is not the case in the other works in literature.

Table 9: Minimum, maximum, and mean differences in accuracy corresponding to features with subsequent ranks in the one round of RFE, and IFE.

|     | Minimum | Maximum | Mean  |
|-----|---------|---------|-------|
| **RFE** | 0.01%   | 0.05%   | 0.03% |
| **IFE** | 0.06%   | 2.72%   | 0.68% |

Table 10: Performance of the proposed system, under different feature combinations.

| **Features**           | **Accuracy** |
|------------------------|--------------|
| MFCC + Top 3 Features  | 0.9946       |
| MFCC + Least 3 Features| 0.9904       |
| MFCC + All features    | 0.9933       |

In line with these findings, as a final evaluation step, the prosodic features are combined with MFCC in a feature combination experiment. For the feature combination experiment, the top three (from ablation experiments), the least three, and all ten handcrafted features, are combined with MFCC in three experiments, and their results are shown in Table 10. From the Table it can be seen that MFCC combined with the top three handcrafted features determined from the feature ablation study produces the best results across all experiments discussed so far (0.9946). This is better than if all handcrafted features are combined with MFCC features (0.9933). In [56], a similar observation is reported, where on-par or higher unweighted average recall (UAR) for emotion classification is achieved when using a subset of the full feature set, specifically, when using 30 out of 88 eGeMAPs features [57], and 100 out of 988 emobase features which is based on the INTERSPEECH 2010 Paralinguistics Challenge featureset [58]. Furthermore, in the current work, when the MFCC features are combined with the least three features, it still performs better than when only MFCC is used (0.9904 vs 0.9895).

Hence, from these results we can conclude that a combination of handcrafted and MFCC features offers the best results. This is in line with findings in literature where, when MFCC and some other features are combined, they offer better performance [14, 15, 16, 32, 39]. This can be attributed to the fact that the MFCC and the other features complement each other. Furthermore, as detailed in section. 3.1.4, spectral features perform well in favorable acoustic conditions, and prosodic features are relatively less affected by channel variations and noise. Hence, a combination of spectral and prosodic features seems ideal. Finally, when considering the proposed best featureset (MFCC + three best handcrafted features), a definite pattern can be established. It is known that a speech signal can be described by its source and system information. In the best featureset which comprises the $F_0$ and HNR that describe the source, the MFCC that describe the system information, and energy that describe the amplitude dynamics of the speech signal, all aspects of the speech signal are addressed.

# 8  Conclusions

Human-computer interactions in Tamil currently takes place mainly in its literary form. However, if the aim is to provide comfort for the common man, then colloquial Tamil must be addressed, and if the prestigious status and heritage of Tamil should be preserved, literary Tamil must still be addressed. To aid in this process, a front-end literary and colloquial Tamil dialect identification system is the first step. To achieve this, in the current work, a 1DCNN-based system trained on a set of handcrafted and MFCC features, is proposed. The handcrafted features are categorized into prosodic, spectral, voice quality and temporal features, and are meant to address the various aspects of speech. The CNN captures the trend of these features over time. The analysis experiment of each of these handcrafted features using parallel LT and CT utterances show that they capture distinguishing characteristics between them. Contradictory to existing LID and DID systems based on prosodic features, the system built with the ten handcrafted features in the current work, performed on par with the MFCC features. This success can be attributed to the proposed 1DCNN system which captures the dynamic trend in these features over time. Furthermore, a feature ablation study to rank each of the ten handcrafted features is carried out. Since the results of feature ablation through recursive feature elimination was ambiguous, an alternate independent feature evaluation method is proposed, to find the rank of each feature. Finally, the quintessential MFCC features are combined with the best three, least three, and all the handcrafted features in a feature combination experiment. The results of this feature combination experiment suggests that: (i) a subset of the handcrafted features offers better performance than using all the handcrafted features, when in combination with MFCC, and (ii) addressing the different aspects of speech, namely the source (through the handcrafted features) and the system (through the MFCC features) can offer the best performance. The authors hope that these proposals are a good initiative towards the progress of human-computer interaction in the user's native language, affirming both the user's preferences and the language's welfare.